\documentclass[11pt]{article}
\pdfoutput=1
\usepackage{fullpage}
\usepackage{xcolor}
\usepackage{cite}
\usepackage{graphicx}
\usepackage{tensor}
\usepackage{amsmath}
\usepackage{amssymb}
\usepackage{subcaption}
\usepackage{diagbox}
\usepackage[utf8x]{inputenc} 
\title{}
\date{}
\renewcommand{\vec}[1]{\mbox{\boldmath$ #1 $}}

\def\beq{\begin{equation}}
\def\eeq{\end{equation}}
\DeclareMathOperator{\artanh}{artanh}
\newcommand*{\cVVuS}{\ensuremath{c_{VV}^{(1),u}}}
\newcommand*{\cVAuS}{\ensuremath{c_{VA}^{(1),u}}}
\newcommand*{\cAVuS}{\ensuremath{c_{AV}^{(1),u}}}
\newcommand*{\cAAuS}{\ensuremath{c_{AA}^{(1),u}}}
\newcommand*{\cVVuO}{\ensuremath{c_{VV}^{(8),u}}}
\newcommand*{\cVAuO}{\ensuremath{c_{VA}^{(8),u}}}
\newcommand*{\cAVuO}{\ensuremath{c_{AV}^{(8),u}}}
\newcommand*{\cAAuO}{\ensuremath{c_{AA}^{(8),u}}}

\allowdisplaybreaks
\begin{document}
\bibliographystyle{utphys}

\newcommand{\be}{\begin{equation}}
\newcommand{\ee}{\end{equation}}
\newcommand\n[1]{\textcolor{red}{(#1)}} 
\newcommand{\diff}{\mathop{}\!\mathrm{d}}
\newcommand{\lb}{\left}
\newcommand{\rb}{\right}
\newcommand{\f}{\frac}
\newcommand{\pd}{\partial}
\newcommand{\tr}{\text{tr}}
\newcommand{\fdiff}{\mathcal{D}}
\newcommand{\im}{\text{im}}
\let\caron\v
\renewcommand{\v}{\mathbf}
\newcommand{\T}{\tensor}
\newcommand{\R}{\mathbb{R}}
\newcommand{\C}{\mathbb{C}}
\newcommand{\Z}{\mathbb{Z}}
\newcommand{\msbar}{\ensuremath{\overline{\text{MS}}}}
\newcommand{\DIS}{\ensuremath{\text{DIS}}}
\newcommand{\abar}{\ensuremath{\bar{\alpha}_S}}
\newcommand{\bb}{\ensuremath{\bar{\beta}_0}}
\newcommand{\rc}{\ensuremath{r_{\text{cut}}}}
\newcommand{\Nd}{\ensuremath{N_{\text{d.o.f.}}}}
\newcommand{\red}[1]{{\color{red} #1}}
\newcommand{\mf}[1]{\mathfrak{#1}}
\newcommand{\cl}[1]{\mathcal{#1}}
\renewcommand{\[}{\begin{equation}\begin{aligned}}
\renewcommand{\]}{\end{aligned}\end{equation}}
\titlepage

\vspace*{0.5cm}

\begin{center}
{\bf \Large Probing new physics in the top sector using quantum information}


\vspace*{1cm} \textsc{
  Rafael Aoude$^a$\footnote{rafael.aoude@ed.ac.uk}, Hannah Banks$^{b}$\footnote{hannah.banks@nyu.edu}, Chris
  D. White$^c$\footnote{christopher.white@qmul.ac.uk} and Martin
  J. White$^d$\footnote{martin.white@adelaide.edu.au}} \\

\vspace*{0.5cm} $^a$ Higgs Centre for Theoretical Physics, School of
Physics and Astronomy, \\The University of Edinburgh, Edinburgh EH9
3JZ, Scotland, UK\\

\vspace*{0.5cm} $^b$ DAMTP, University of Cambridge, Wilberforce Road,
Cambridge, CB3 0WA, UK\\

\vspace*{0.5cm} $^c$ Centre for Theoretical Physics, School of Physical and Chemical Sciences, \\
Queen Mary University of London, 327 Mile End
Road, London E1 4NS, UK\\

\vspace*{0.5cm} $^d$ ARC Centre of Excellence for Dark Matter Particle
Physics \& CSSM,  \\ Department of Physics, University of Adelaide,
Adelaide, SA 5005, Australia\\

\end{center}

\vspace*{0.5cm}

\begin{abstract}
Recent studies have shown that quantitative concepts from quantum information  theory can play a role in analysing collider physics, including elucidating new physics. In this paper, we study various QI measures including {\it magic}, {\it trace distance} and {\it fidelity distance}, in generic new physics scenarios modelled by the Standard Model Effective Field Theory. We argue that such measures can indeed show up differences with respect to the pure Standard Model, and we compare our results with similar findings for the {\it concurrence} discussed previously in the literature. We examine the relative sensitivity of different measures to new physics in two-dimensional bins of the top pair invariant mass and scattering angle, finding that the concurrence, magic and trace distance each emerge as the best measure in at least some regions of the phase space. This highlights the importance of exploring multiple quantum information measures in the hunt for beyond the Standard Model physics. 
\end{abstract}

\vspace*{0.5cm}

\newpage

\section{Introduction}
\label{sec:intro}

In recent years, a growing body of work has examined the use of particle collider experiments to perform fundamental tests of quantum theory. The currently operating Large Hadron Collider (LHC) offers unprecedentedly high energies at which to verify concepts such as
quantum entanglement, which was first proposed for top quark spins in
refs.~\cite{Afik:2020onf,Afik:2022kwm} 
(see refs.~\cite{Dong:2023xiw,Aoude:2022imd,Fabbrichesi:2021npl,Severi:2021cnj,Afik:2022kwm,Aoude:2022imd,Aguilar-Saavedra:2022uye,Fabbrichesi:2022ovb,Afik:2022dgh,Severi:2022qjy,Aguilar-Saavedra:2023hss,Han:2023fci,Simpson:2024hbr,Aguilar-Saavedra:2024hwd,Maltoni:2024csn}
for follow-up work, and ref.~\cite{Barr:2024djo} for a recent comprehensive and pedagogical review). Experimental studies from both the ATLAS and CMS collaborations can be found in
refs.~\cite{ATLAS:2023jzs,ATLAS:2023fsd,CMS:2024hgo}.

Given this context, one can ask if other ideas from quantum computation / information theory may prove useful in a collider context. To this end, ref.~\cite{White:2024nuc} examined a quantity
known as {\it magic} which, roughly speaking, classifies whether or not quantum states possess a genuine computational advantage over their
classical counterparts~\cite{Nielsen:2012yss}. Magic has been widely studied in quantum
systems~\cite{Beverland_2020,PhysRevApplied.19.034052,Leone:2023avk,Qassim2021improvedupperbounds,Leone:2021rzd,Haug:2023ffp,Magic1,PhysRevA.108.042408,Gu:2023qqq,Tirrito:2023fnw,Turkeshi:2023lqu,Leone:2024lfr},
not least due to its pivotal role in the design of potential fault-tolerant quantum computers. Other applications of magic in high energy physics and related quantum systems can be found in
refs.~\cite{Tarabunga:2023ggd,Frau:2024qmf,Lami:2024osd,Robin:2024bdz}. How to produce and manipulate magic quantum states very much remains an open question, and thus one of the principal aims of ref.~\cite{White:2024nuc} was to showcase a particular high energy quantum system -- the same top pair production process utilised in previous studies of entanglement -- and show that it provides a
natural playground for investigating magic. Furthermore, the amount of magic in the (mixed) top quark final state depends upon the kinematic properties of the top quarks (their velocity and scattering angle), and is thus tunable via event selection. An additional hope raised by
ref.~\cite{White:2024nuc} is that quantifying magic may itself prove useful in distinguishing new physics theories from the Standard Model
of particle physics, and the aim of this paper is to make this concrete. 

We note that the study of magic in collider processes has also been the subject of other recent works. In particular, ref.~\cite{Liu:2025frx} placed bounds on magic in two-qubit systems, where the latter naturally arise in a collider setting. Ref.~\cite{Liu:2025qfl} studied how magic is produced in scattering processes in Quantum Electrodynamics (QED), and ref.~\cite{Fabbrichesi:2025ywl} quantified magic in various processes other than the top quark pair production process considered in ref.~\cite{White:2024nuc}. It also studied other useful measures from quantum information (QI) theory, such as the {\it trace distance}, {\it fidelity distance} and {\it concurrence}  that we will review below (see also refs.~\cite{Barr:2022wyq,AshbyPickering:2022umy,Aguilar-Saavedra:2022wam,Aguilar-Saavedra:2022mpg,Fabbri:2023ncz,Aoude:2023hxv,Fabbrichesi:2023cev,Sakurai:2023nsc,Altomonte:2023mug,Afik:2024uif,Aguilar-Saavedra:2024vpd,Aguilar-Saavedra:2024whi,Grabarczyk:2024wnk,Morales:2024jhj,Altomonte:2024upf,Han:2024ugl,Cheng:2024rxi,Subba:2024aut,Subba:2024mnl,Horodecki:2025tpn,DelGratta:2025qyp,Nason:2025hix,Grossi:2024jae,Cheng:2025cuv} for further work in this area). Comparisons were made between these different measures as possible probes of new physics, and our present study is directly inspired by this approach. Excitingly, the CMS collaboration has now produced a measurement of magic in top quark pair production~\cite{CMS:2025cim}, and there are widespread international efforts to coordinate the community of researchers working on quantum information applied to collider physics~\cite{Afik:2025ejh}.

In describing new physics, we will consider a relatively agnostic approach to parametrising beyond the Standard Model (BSM) physics, given by the well-known approach of effective field theory (EFT) ~\cite{Weinberg:1978kz}. There, one regards the Standard Model (SM) as the lowest-order contribution to a higher theory, in which new physics is associated with a particular energy scale, $\Lambda$. One may then supplement the SM Lagrangian density with a series of effective
operators built from SM fields and their derivatives, each
successively suppressed by higher inverse powers of $\Lambda$, so that the operators themselves have ever increasing mass dimension. At any given dimension, there is a finite, irreducible set of operators permitted by gauge
invariance. Any given complete theory of BSM physics would in principle uniquely fix the effective operators observed at lower energies. However, leaving the coefficients of the effective operators (known as {\it Wilson coefficients}) undetermined amounts to being entirely agnostic about the nature of the new physics. By considering suitable observables, one may then fit the
undetermined Wilson coefficients to data, where a non-zero result for any such coefficient amounts to the discovery of new physics.

Given that the SM Lagrangian density (in four spacetime dimensions) has mass dimension 4, the first effective operators appear at dimension 5. In fact there is a single such operator, the so-called Weinberg operator, associated with non-zero neutrino masses. A wider spectrum of 59 independent operators appears at dimension six~\cite{Buchmuller:1985jz,BURGES1983464,Leung:1984ni,Grzadkowski:2010es}, where the choice of basis for these operators is not unique. Fits to data using various bases have been made in e.g. refs.~\cite{Giani:2023gfq,Hartland:2019bjb,Brivio:2019ius,Biekotter:2018ohn,Ellis:2018gqa,daSilvaAlmeida:2018iqo,Aebischer:2018iyb,Ellis:2020unq,Bruggisser:2021duo,Bruggisser:2022rhb,Ethier:2021bye,Buckley:2015lku,Buckley:2015nca,Ethier:2021ydt}, and we will here adopt the choice of
ref.~\cite{Aguilar-Saavedra:2018ksv}. As a companion study to ref.~\cite{White:2024nuc}, we will explicitly demonstrate how specific dimension six operators change the magic of top quark pairs at the LHC, and our general conclusion will be that magic tends to increase. The latter is not surprising: magic vanishes for states that have a particularly simple Pauli spectrum (to be reviewed in more detail below). This typically happens for non-entangled states, and for those which are maximally entangled. It was noted in ref.~\cite{Aoude:2022imd} that SMEFT operators tend to decrease the entanglement of top quark pairs and this was later confirmed at higher orders in ref.~\cite{Severi:2022qjy}. This in turn makes it plausible -- but not automatic -- that the magic should increase. There are kinematic regions, however, where the magic can decrease, and it is also true that the pattern of magic variation can be different for different operators. As noted above, this opens up the possibility of using magic as a useful observable to distinguish different new physics scenarios, in a way that provides links with other systems studied in QI theory. Given that magic provides complementary information to entanglement, our results provide a highly useful accompaniment to those of ref.~\cite{Aoude:2022imd}. 

The magic is an intrinsic property of a single quantum density matrix, and as such may differ between density matrices obtained from different theories (e.g. the pure SM and the SMEFT). The QI measures considered in ref.~\cite{Fabbrichesi:2025ywl} -- specifically the trace distance and fidelity distance -- are novel in that their definition explicitly involves a pair of density matrices, so that the observables are intentionally designed to measure the `closeness' of two (mixed) quantum states in some prescribed way. One might then hope that such measures provide a more sensitive probe of new physics than measures which are already non-zero in the SM, and ref.~\cite{Fabbrichesi:2025ywl} applied this idea in the context of constraining anomalous couplings of the $\tau$ lepton. The authors found that the trace distance typically outperforms other observables, but that does not mean that this will necessarily be true in other processes. In what follows, we will present results for the variety of QI measures discussed above, as applied to top quark pair production in the SMEFT and the SM, alongside the concurrence which has been previously considered in ref.~\cite{Aoude:2022imd}. We study the relative sensitivity of these measures in two-dimensional bins in the parameter space of top quark invariant mass and scattering angle, finding that no single measure outperforms all others across the board (including the raw Fano coefficients themselves). Thus, it is important to consider a range of QI measures in constraining new physics, where the most sensitive observable depends both on the collider process being considered, and the kinematic region being probed.
Our results will prove useful for further studies in this area, and will also serve to strengthen the already prevalent dialogue between QI and collider physics~\cite{Afik:2025ejh}.

We begin in section~\ref{sec:review}, by briefly reviewing the concepts necessary for the ensuing discussion. In section~\ref{sec:magic}, we then show and interpret results for the magic stemming from different SMEFT operators, at both parton and proton level. In section~\ref{sec:QI}, we 
examine the concurrence, trace distance and fidelity distance for top quark pair production, and compare their use in probing new physics to the magic. We discuss our results
and conclude in section~\ref{sec:conclude}.

\section{Review of necessary concepts}
\label{sec:review}

In this section, we review a number of salient details regarding the SMEFT, and also the various quantum information measures that we will use in subsequent sections. A more detailed review of magic in the present context may be found in ref.~\cite{White:2024nuc}.

\subsection{Effective Field Theory and the SMEFT}
\label{sec:SMEFT}
The framework of effective field theory can be used whenever the scale $\Lambda$ associated with new physics (e.g. the lowest new particle mass) is much higher than the energy of a given experiment. One may then supplement the SM Lagrangian density ${\cal L}_{\rm SM}$ with a series of gauge-invariant higher-dimensional operators composed of SM fields, so that the full Lagrangian under consideration has the schematic form
\begin{equation}
  {\cal L}={\cal L}_{\rm SM}+\sum_{n=1}^\infty \frac{1}{\Lambda^{n}}
  \sum_i c^{(n)}_i{\cal O}^{(n)}_i~.
  \label{Ltot}
\end{equation}
Here the first sum is over increasing inverse powers of the new physics scale $\Lambda$, and the second sum includes all possible operators ${\cal O}^{(n)}_i$ at a given power which, if the coefficients $c_i^{(n)}$ are dimensionless, must have mass dimension $[{\cal O}_i^{(n)}]=4+n$. A given new physics theory will fix the values of the coefficients $\{c_i^{(n)}\}$, such that at energy scales $E\sim\Lambda$, one must resum the effective field theory expansion, given that all terms will be equally important. At energy scales $E\ll\Lambda$, the series in $n$ is highly convergent, and is typically truncated at $n=2$ (dimension six). The $n=1$ term can be ignored for our purposes: it is associated with non-zero neutrino masses, and will be irrelevant for top quark pair production processes.

At a given mass dimension, there is a finite number of possible operators, which are not unique given that one may use the equations of motion to redefine the basis. Here, we will adopt the basis of ref.~\cite{Aguilar-Saavedra:2018ksv}, which is particularly well-suited to new physics corrections in the top quark sector, and is itself based on the seminal analysis of ref.~\cite{Grzadkowski:2010es}. Of the 59 possible operators at dimension six, only a subset contribute to top quark pair production. To reduce their number, we make the same simplifying assumptions as ref.~\cite{Aguilar-Saavedra:2018ksv}, namely that of a
unit CKM matrix, and retaining only the third generation Yukawa couplings (see section~4.1 of that reference). Then the list of contributing operators comprises first those involving the gluon:
\begin{align}
  {\cal O}_G=g_s f^{ABC}&G_\nu^{A,\mu}\,G_{\rho}^{B,\nu}\,G_\mu^{C,\rho}~,\quad
  {\cal O}_{\phi G}=\left(\phi^\dag\phi-\frac{v^2}{2}\right)G^{\mu\nu}_A\,
  G^A_{\mu\nu}~,\notag\\
  &{\cal O}_{tG}=g_s \left(\bar{Q}\sigma^{\mu\nu} T^A t
  \right)\tilde{\phi}\,G^A_{\mu\nu}+{\rm h.c.}
  \label{O1}
\end{align}
Of these operators, the triple gluon coupling $c_G$ is the most highly constrained currently, due e.g. to the possibility of probing it  in multijet processes and  top pair production~\cite{Englert:2018byk,Buckley:2015nca,Buckley:2015lku,Giani:2023gfq,Ethier:2021bye,Hartland:2019bjb}.

In these expressions capital letters denote adjoint (colour) indices; $G_{\mu\nu}^A$ is the gluon field strength; $\phi$ the SM Higgs doublet with charge conjugate $\tilde{\phi}=i\sigma^2\phi^*$ (in terms of the Pauli matrix
$\sigma^2$); $v$ the Higgs vacuum expectation value (VEV); $Q$ the doublet of SM heavy quark fields; $T^A$ an SU(3) colour generator with associated structure constants $f^{ABC}$; and $t$ the top quark field. To eq.~(\ref{O1}) we must add various 4-fermion operators involving the top quark, subject to the simplifying assumptions mentioned above. One may distinguish the colour octet operators
\begin{align}
\label{O2}
    {\cal O}_{Qq}^{(8,1)}&=\left(\bar{Q}_L\gamma_\mu T^A Q_L\right)
    \left(\bar{q}_L\gamma^\mu T^A Q_L\right)~,\quad
    {\cal O}_{Qq}^{(8,3)}=\left(\bar{Q}_L\gamma_\mu T^A\tau^a Q_L\right)
    \left(\bar{q}_L\gamma^\mu T^A\tau^a q_L\right)~,\notag\\
    {\cal O}_{tu}^{(8)}&=\left(\bar{t}_R\gamma_\mu T^A t_R\right)
    \left(\bar{u}_R\gamma^\mu T^A u_R\right)~,\quad
    {\cal O}_{td}^{(8)}=\left(\bar{t}_R\gamma_\mu T^A t_R\right)
    \left(\bar{d}_R\gamma^\mu T^A d_R\right)~,\notag\\
    {\cal O}_{Qu}^{(8)}&=\left(\bar{Q}_L\gamma_\mu
    T^A Q_L\right)\left(\bar{u}_R\gamma^\mu T^A u_R\right)~,
    \quad {\cal O}_{Qd}^{(8)}=\left(\bar{Q}_L\gamma_\mu
    T^A Q_L\right)\left(\bar{d}_R\gamma^\mu T^Ad_R\right)~,\notag\\
    &\qquad\qquad\qquad\qquad {\cal O}_{tq}^{(8)}=\left(\bar{t}_R\gamma_\mu T^A t_R\right)
    \left(\bar{q}_L\gamma^\mu T^A q_L\right)~,
\end{align}
from their colour singlet counterparts $\{{\cal O}^{(1,1)}_{Qq},{\cal
  O}^{(1,3)}_{Qq},{\cal O}_{ij}^{(1)}\}$, which are  obtained from analogous formulae with  the SU(3) generators removed. In these expressions we have further used $Q_L$ to denote the SU(2) doublet of the left-handed heavy quark fields, and $R$ to denote a right-handed singlet. Finally, $\tau^a$ denotes an SU(2) generator. Taking into account both octet and singlet operators and 
combining with eq.~(\ref{O1}) gives a list of 17 operators, whose Wilson coefficients can be labelled similarly to the operators themselves (e.g. $c_G$, $c^{(8)}_{td}$). We will furthermore define the normalised coefficients
\begin{equation}
    \bar{c}_I=\frac{c_I}{\Lambda^2}~,
    \label{cbardef}
\end{equation}
which have the cut-off scale divided out, given that it is only this combination that is directly constrained by experiment. Note also that not all operators in eq.~(\ref{O2}) are constrained individually in top quark pair production at the LHC. Rather, only specific linear combinations are constrainable, collected here in eq.~(\ref{cVV}). Current bounds on relevant Wilson coefficients can be found in refs.~\cite{Celada:2024mcf,terHoeve:2025gey,Cornet-Gomez:2025jot}. While quantum information measures can be used to constrain such coefficients in a global fit, as explored in~\cite{Cornet-Gomez:2025jot}, the values used in the following sections will be chosen for visualisation purposes.

\subsection{Top pair production and the density matrix}
\label{sec:tpp}

The most common production mode for top quarks at the LHC is that of {\it pair production}, in which a top quark is produced alongside an anti-top. Quantum field theory (QFT) tells us that all possible final states will be produced, weighted by appropriate probabilities. We then say that the (anti-)top pair is in a {\it mixed state}, rather than a
uniquely predicted {\it pure state}. Given that top (anti-)quarks are spin-1/2 particles, the top pair forms a two-qubit system in the language of quantum information theory. The particular mixed spin state of the top pair will depend upon the kinematics (e.g. the top quark velocity and scattering angle), and may include configurations in which the top spins are entangled by quantum effects.

As detailed in refs.~\cite{Afik:2020onf,Afik:2022kwm}, we can describe a general mixed state of top quarks using the well-known density matrix formalism. Let $|I\lambda\rangle$ denote a particular initial
state with particle content $I$ (e.g. $gg$, $u\bar{u}$, etc.) with associated quantum numbers $\lambda$ (e.g. denoting colours and spins). If ${\cal A}_{\alpha\beta}^{I\lambda}(M,\vec{k})$ is the scattering amplitude from the state $|I\lambda\rangle$ to a final state with top pair invariant mass
$M$, top quark unit direction vector $\hat{\vec{k}}$, and (anti-)top spin indices $\alpha$ ($\beta$), then we may define the so-called $R$-matrix for this process as 
\begin{equation}
  R^{I\lambda}_{\alpha\beta,\alpha'\beta'}(M,\hat{k})=
  {\cal A}_{\alpha\beta}^{I\lambda}(M,\hat{\vec{k}})
  \left[{\cal A}_{\alpha'\beta'}^{I\lambda}(M,\hat{\vec{k}})\right]^\dag~.
  \label{RIlamdef}
\end{equation}
This can then be averaged over $\lambda$ to produce the {\it production (spin) density matrix} for the channel $I$
\begin{equation}
  R^I_{\alpha\beta,\alpha'\beta'}(M,\hat{\vec{k}})=\frac{1}{N_\lambda}
  \sum_\lambda R^{I\lambda}_{\alpha\beta,\alpha'\beta'}(M,\hat{\vec{k}})~.
  \label{RIdef}
\end{equation}
Using this, it is convenient to define a normalised production density matrix with unit trace, via
\begin{equation}
  \rho^I=\frac{R^I}{{\rm Tr}(R^I)}~.
  \label{rhoIdef}
\end{equation}

Given that one does not always consider final states fully differentially in kinematics in experiment, it is also convenient to consider the R matrix averaged over the angular directions of the final state top quark:
\begin{equation}
 \bar{R}^I = (4\pi)^{-1} \int d \Omega R^I(\hat{s},\bf{k})~,
 \label{Rav}
\end{equation}
where $d\Omega$ denotes the element of solid angle. From this one may then construct an angularly averaged normalised production density matrix $\bar{\rho}^I$ similarly to equation~\ref{rhoIdef}. 


Returning to the fully differential $R$-matrix of eq.~(\ref{RIdef}),
this can be generically decomposed as
\begin{equation}
  R^I=\tilde{A}^I I_4+\sum_i\left(\tilde{B}^{I+}_i \sigma_i\otimes I_2
  +\tilde{B}^{I-}_i I_2\otimes\sigma_i\right)+\sum_{i,j}\tilde{C}^I_{ij}\sigma_i
  \otimes\sigma_j~,
  \label{Rdecomp}
\end{equation}
where $I_n$ is an $n$-dimensional identity matrix, and $\sigma_i$ are the Pauli matrices. The various quantities $\{\tilde{A}^I,\tilde{B}^{I\pm}, \tilde{C}^I_{ij}\}$ are known as {\it Fano coefficients}, and will depend in this case on the kinematics of the final state. Substitution of 
eq.~(\ref{Rdecomp}) into eq.~(\ref{rhoIdef}) yields 
\begin{equation}
  \rho^I=\frac{R^I}{4\tilde{A}^I}~,
  \label{rhoIdecomp}
\end{equation}
which thus has a form similar to eq.~(\ref{Rdecomp}),
but with normalised Fano coefficients
\begin{equation}
  B^{I\pm}_i=\frac{\tilde{B}^{I\pm}_i}{\tilde{A}^I},\quad C^I_{ij}=
  \frac{\tilde{C}^I_{ij}}{\tilde{A}^I}~.
  \label{BCnorm}
\end{equation}
Physically, the coefficient $\tilde{A}^I$ in a given partonic channel $I$ is related to the overall cross-section, and $\tilde{B}^{I\pm}$ measures the polarisation of each individual top (anti-)quark. The coefficients $\tilde{C}^I_{ij}$ quantify the spin correlations between
the two top particles, and are thus the relevant  quantities for discussing quantum entanglement. In order to fully fix the Fano coefficients, we must define a basis, corresponding to aligning the $(1,2,3)$ directions in spin space with a given set of orthonormal spatial vectors. A common choice is the {\it helicity basis}, in which one defines the 3-vectors\footnote{Note that some references make the definition $ \hat{\vec{n}}=\hat{\vec{r}}\times \hat{\vec{k}}$. Our choice was made to match that of ref.~\cite{Aoude:2022imd} which we rely upon for expressions for the SMEFT contributions to the Fano coefficients.}
\begin{equation}
  \hat{\vec{r}}=\frac{\hat{\vec{p}}-\cos\theta\,\hat{\vec{k}}}{\sin(\theta)},\quad
  \hat{\vec{n}}=\hat{\vec{k}}\times \hat{\vec{r}}~,
  \label{rndef}
\end{equation}
in terms of the first incoming beam direction $\hat{\vec{p}}$ and scattering angle $\cos\theta=\hat{\vec{k}}\cdot \hat{\vec{p}}$.
Since this is a 2-to-2 scattering process, the kinematics is fully defined by two variables. We will mostly use the top quark velocity, and the cosine of the scattering angle:
\begin{equation}
    \beta = \sqrt{1-\frac{4m_t^2}{\hat{s}}}~, \qquad z= \cos\theta~.
\end{equation}

At
leading order (LO) in QCD, the top polarisations vanish, as do certain components of the spin correlation matrix:
\begin{equation}
  \tilde{B}_i^{I\pm}=\tilde{C}^I_{nr}=\tilde{C}^I_{nk}=0~.
  \label{Bzero}
\end{equation}
Furthermore, the remaining spin correlation coefficients are symmetric:
\begin{equation}
  \tilde{C}^I_{ij}=\tilde{C}^I_{ji}~.
  \label{Csym}
\end{equation}
The various properties listed here are potentially modified by BSM contributions, and it is thus useful to keep them in mind in what follows.

The above Fano coefficients are defined at the level of individual partonic channels, $I$. For fully realistic proton initial states, one must combine these, weighting by the appropriate parton luminosity function 
    \begin{equation}
    L^I (M) = \frac{1}{s}\int_\tau^1 \frac{dx}{x} f_a (x,\sqrt{\hat{s}})f_b (\tau/x,\sqrt{\hat{s}})~. 
    \end{equation}
Here $a$, $b$ denote the two partons in the partonic channel $I$, $f_a(x,Q)$ is the parton distribution function for parton $a$ with momentum fraction $x$ and factorisation scale $Q$, and $\sqrt{\hat{s}}$ is the partonic centre of mass energy, which is in turn equal to the top quark invariant mass $M$ at leading order. The parameter $\tau$ is defined as 
\begin{equation}
\tau\equiv\frac{\hat{s}}{s}~,
\end{equation}
where $\sqrt{s}$ is the hadronic centre-of-mass energy of 13 TeV. The averaged Fano coefficients then take the form e.g. 
\begin{equation}
    C^{\rm pp}_{ij}=\sum_I 
    \frac{w_I(z,\beta) \tilde{C}^{I}_{ij}}{\tilde{A}^I},
    \quad w_I=\frac{L^I(\beta)\tilde{A}^I(z,\beta)}
    {\sum_J L^J(\beta)\tilde{A}^J(z,\beta)}~.
    \label{Cpp}
\end{equation}
It should be emphasised that for initial states involving a pair of quarks, the (anti)-quark could come from either proton and both cases, which we denote via $q \bar{q}$ and $ \bar{q} q$, need to be explicitly accounted for in the sum. As discussed in appendix ~\ref{app:FanoSM}, when working fully differentially in the helicity basis, the Fano coefficients for the $q \bar{q}$ and $ \bar{q} q$ processes are different as a result of interchanging the directions in which the quark and the anti-quark are travelling. It is thus essential to treat these two cases as distinct channels in the sum, albeit weighted by the same luminosity function, $L^{q\bar{q}} =L^{\bar{q} q} $. When working with the angularly averaged Fano coefficients however, the expressions for the $q \bar{q}$ and $ \bar{q} q$ processes are identical. As such, many works combine these into a single channel for the purpose of computing the sum, working with a redefined quark luminosity function  $\tilde{L}^{q\bar{q}} =  L^{q\bar{q}} +  L^{\bar{q} q} = 2L^{q\bar{q}} $ which already symmetrises over the $q\bar{q}$ and $\bar{q}q$ channels. 

Throughout, we use the \texttt{NNPDF4.0} NNLO PDF set~\cite{NNPDF:2021njg}, as implemented in the \texttt{parton} Python package, based on the \texttt{LHAPDF} project~\cite{Buckley:2014ana}. The factorisation scale is set to $\mu_F=\sqrt{\hat{s}}$, and strong coupling constant $\alpha_s=0.118$.

\subsection{Magic in bipartite qubit systems}
\label{sec:topmagic}
Given a pure quantum state $|\psi\rangle$ of an $n$-qubit system, we can consider its {\it Pauli spectrum}
\begin{equation}
  {\rm spec}(|\psi\rangle)=\{\langle\psi|P|\psi\rangle,\quad P\in
  {\cal P}_n\}~,
  \label{Paulispec}
\end{equation}
namely the set of expectation values of {\it Pauli strings}
\begin{equation}
  {\cal P}_n=P_1\otimes P_2\otimes \ldots\otimes P_N,\quad
  P_a\in\{\sigma^{(a)}_1,\sigma^{(a)}_2,\sigma_3^{(a)},I^{(a)}\}~,
  \label{Paulistring}
\end{equation}
composed of Pauli or identity matrices acting on each individual qubit $a$. The Pauli spectrum is useful in that there is a set of states known as {\it stabiliser states} whose Pauli spectrum is particularly simple. It consists of $2^n$ expectation values equal to $\pm1$ and the rest
zero, and such states can be made in quantum computers by operating on the state
\begin{displaymath}
|0\rangle\otimes |0\rangle\otimes\ldots\otimes|0\rangle~,
\end{displaymath}
with particular quantum gates\footnote{The particular quantum gates are the CNOT, Hadamard and phase gates. These details will not be necessary in this paper.}. The importance of stabiliser states for quantum computation stems from the Gottesman-Knill theorem which, roughly speaking, says that quantum circuits featuring only stabiliser states offer no computational advantage over classical computers. Stabiliser states can include maximally entangled states, and thus this tells us that entanglement by itself is not the be all and end all of quantum computation. It is then useful to have
quantities that efficiently characterise the Pauli spectrum of a given quantum state, and which can encode its ``non-stabiliserness''. The word {\it magic} has been adopted in recent years to refer to non-stabiliserness, and various concrete definitions have been proposed and studied~\cite{Beverland_2020,PhysRevApplied.19.034052,Leone:2023avk,Qassim2021improvedupperbounds,Leone:2021rzd,Haug:2023ffp,Magic1,PhysRevA.108.042408,Gu:2023qqq,Tirrito:2023fnw,Turkeshi:2023lqu,Leone:2024lfr}. Magic is crucial in the development of fault-tolerant quantum computation
algorithms, thus how to produce and manipulate magic are highly topical research questions.

Motivated by developing research links between quantum information theory and high energy physics, ref.~\cite{White:2024nuc} examined the role of magic in top pair production. Given its status as a 2-qubit system, the top pair final state provides a natural playground for studying magic, which also allows to study how magic can be enhanced by going to certain kinematic regions. This may in turn be useful for examining magic in other quantum systems, but it could also be (as mentioned above) that magic is a good observable for distinguishing new  physics from the SM. As its definition of magic, ref.~\cite{White:2024nuc} adopted the so-called {\it Second Stabiliser
  Renyi entropy (SSRE)}, given (for a mixed quantum state with density matrix $\rho$) by
\begin{equation}
  M_2(\rho)=-\log_2\left(\frac{\sum_{P\in{\cal P}_n}{\rm Tr}^4(\rho P)}
  {\sum_{P\in{\cal P}_n}{\rm Tr}^2(\rho P)}\right)~.
  \label{M2tildedef}
\end{equation}
Substituting eqs.~(\ref{Rdecomp}, \ref{rhoIdecomp}) yields
\begin{equation}
  M_2(\rho^I)=
  -\log_2\left(\frac{1+\sum_i [(B_i^{I+})^4+(B_i^{I-})^4]+\sum_{i,j}(C_{ij}^I)^4}
  {1+\sum_i [(B_i^{I+})^2+(B_i^{I-})^2]+\sum_{i,j}(C_{ij}^I)^2}
  \right)~,
  \label{M2tilderes}
\end{equation}
which we will use repeatedly in what follows.

It is helpful when introducing non-zero Wilson coefficients for SMEFT operators to separate out the contributions to the normalised density matrix that arise from the SM and BSM contributions (which we label as ``EFT'' to follow the notation of ref.\cite{Aoude:2022imd}):
\begin{equation}
\rho^I=\frac{R^{I,{\rm SM}}+R^{I,{\rm EFT}}}{{\rm Tr}(R^{I,{\rm SM}})+{\rm Tr}(R^{I,{\rm EFT}})}~.
\end{equation}
To study the impact of new physics on the magic, we expand equation~\ref{M2tilderes} in the Wilson coefficients, to both linear ($\mathcal{O}(\Lambda^{-2})$) and quadratic ($\mathcal{O}(\Lambda^{-4})$) order.\footnote{For the avoidance of doubt, when we refer to the linear or quadratic magic (or indeed any of the other QI measures studied) we refer to the explicit expansion of the expression to the named order in the Wilson coefficients, as opposed to computing the expression using a density matrix formed from Fano coefficients including SMEFT contributions of that order. } Analytic expressions for the SMEFT contributions to the Fano coefficients in the helicity basis (as defined in equation~\ref{rndef}) were obtained in ref.~\cite{Aoude:2022imd}, and are collected here in appendix~\ref{app:Fano}. In appendix~\ref{app:FanoSM} we present expressions for the pure SM contributions to the Fano coefficients in this basis, working with the appropriate normalisation to consistently combine them with the SMEFT contributions. Expressions for the angularly averaged Fano coefficients, working now in the $\textit{beam}$ basis, are provided in Appendix ~\ref{app:ang}, taken again from ref.~\cite{Aoude:2022imd}.

\subsection{Other QI measures}
\label{sec:tracedistance}

As already discussed in the introduction, magic is not the only QI observable that is relevant for collider physics applications. Another commonly analysed criterion is the {\it concurrence}~\cite{Wootters:1997id}, which quantifies the degree of spin entanglement of the top pair final state. Given a two-qubit density matrix $\rho$ and its complex conjugate $\rho^*$, one may construct the related matrix
\begin{equation}
\omega=\sqrt{\sqrt{\tilde{\rho}}\rho\sqrt{\tilde{\rho}}}~,\quad
\tilde{\rho}=(\sigma_2\otimes\sigma_2)\rho^*(\sigma_2\otimes\sigma_2)~.
\label{omegadef}
\end{equation}
Denoting the eigenvalues of this matrix (in decreasing order) by $\{\lambda_i\}$, the concurrence is given by
\begin{equation}
    C[\rho]={\rm max}(0,\lambda_1-\lambda_2-\lambda_3-\lambda_4)~.
    \label{Cdef}
\end{equation}
As for the magic, the concurrence is a property of an individual quantum state. That concurrence and magic do not provide the same information is due to the fact that non-stabliserness and entanglement are not purely complementary.

Recently, ref.~\cite{Fabbrichesi:2025ywl} argued that two other measures are also potentially useful, and we review their definitions here (see also the classic ref.~\cite{Nielsen:2012yss} for a textbook treatment). Both of these quantities are designed to measure how `close' a given quantum state (expressed in terms of its density matrix) is to another. This in turn has clear connotations in investigating new physics models: by taking the two density matrices to correspond to the presence and absence of new physics respectively, one obtains a bona fide QI viewpoint on whether quantum states produced at the LHC are close to the SM or not.

To define the measures of interest, let $\rho$ and $\varsigma$ be density matrices corresponding to two different quantum states. Then the {\it trace distance} is defined by
\begin{equation}
    {\cal D}^{T}(\rho,\varsigma)
    = \frac12 || \rho-\sigma||=
    \frac12{\rm Tr}\sqrt{(\rho-\varsigma)^\dag 
    (\rho-\varsigma)}~.
    \label{tracedist}
\end{equation}
This is positive semi-definite ${\cal D}^T\geq 0$, and invariant under unitary transformations of $\rho$ and/or $\varsigma$. As emphasised in refs.~\cite{Fabbrichesi:2025ywl,Nielsen:2012yss}, eq.~(\ref{tracedist}) can be seen as a quantum generalisation of an analogous classical quantity, that itself corresponds to the so-called $L_1$- or {\it Kolmogorov}-distance between two probability distributions. Indeed, the quantum trace distance reduces to its classical counterpart if the two density matrices $\rho$ and $\varsigma$ mutually commute. 
To gain more insight into this measure for the SMEFT, we may take the distance between $\rho_{\rm SM}$ and the SMEFT density matrix up to linear order in the Wilson coefficients ($\rho_{\rm d6}$): 
\begin{align}
\rho_{\rm d6} 
    = \frac{R^{\rm SM} + \frac{1}{\Lambda^2} R^{(\rm d6)}}{\text{tr}[R^{\rm SM}] + \frac{1}{\Lambda^2} \text{tr}[R^{(\rm d6)}]} 
    = \frac{R^{\rm SM} + \frac{1}{\Lambda^2} R^{(\rm d6)}}{\text{tr}[R^{\rm SM}]}\frac{1}{1 + \frac{1}{\Lambda^2} \text{tr}[R^{(\rm d6)}]/\text{tr}[R^{\rm SM}]}.
\end{align}
Since at this order, there's no proliferation of different terms from the density matrix to the observables, we can expand the former directly at $1/\Lambda^2$ order, which gives us
\begin{align}
\rho_{\rm d6}  = \rho_{\rm SM} + \frac{1}{\Lambda^2}\frac{1}{\text{tr}
[R^{\rm SM}]}\left( R^{(\rm d6)} - \text{tr}[R^{(\rm d6)}] \right) + \cdots
\end{align}
Given that our density matrices are hermitian, the trace distance between the linear-SMEFT and the SM is
\begin{equation}
    {\cal D}^{T}(\rho_{\rm d6},\rho_{\rm SM})=
    \frac12{\rm tr}\sqrt{(\rho_{\rm d6} -\rho_{\rm SM})^\dag 
    (\rho_{\rm d6} -\rho_{\rm SM})}
    =
    \frac12\frac{1}{\Lambda^2}\frac{1}{\text{tr}[R^{\rm SM}]}{\rm tr}\left( 
    |R^{(\rm d6)} - \text{tr}[R^{(\rm d6)}]| \right)~.
\end{equation}
Despite na\"{i}ve appearances, the right-hand side is non-zero in general due to the absolute value inside the trace. We also see  that the trace distance is manifestly linear in the Wilson coefficients, with no zeroth-order term.

Our second QI measure involving two different states is the {\it fidelity}:
\begin{equation}
    {\cal F}(\rho,\varsigma)
    ={\rm Tr}\,\sqrt{\sqrt{\rho}\,\,\varsigma\,\,{\sqrt{\rho}}}~.
    \label{fidelity}
\end{equation}
Despite appearances, this is symmetric in the two density matrices i.e. ${\cal F}(\rho,\varsigma)={\cal F}(\varsigma,\rho)$. Furthermore, one may establish the bounds:
\begin{equation}
    0\leq {\cal F}\leq 1~,
    \label{Fbounds}
\end{equation}
where the upper limit is saturated by $\rho=\varsigma$, and the lower limit if $\rho$ and $\varsigma$ have orthogonal support~\cite{Nielsen:2012yss}. 
As for the trace distance, the fidelity is invariant under unitary transformations of either density matrix, and reduces to a classical counterpart if the density matrices mutually commute. Whereas trace distance increases as states become less alike, the fidelity decreases. This motivates the definition of the {\it fidelity distance}
\begin{equation}
{\cal D}^F=\sqrt{1-{\cal F}^2}~,
\label{FD}
\end{equation}
which instead increases from zero as states move away from each other in Hilbert space. Which quantum information measure to use in any given situation depends upon the question being asked. Our motivation here, inspired by ref.~\cite{Fabbrichesi:2025ywl}, is that some QI measures may be more sensitive to new physics than others, which may itself depend on the particular scattering processes or kinematic region being considered. 

We have now introduced all of the QI concepts needed for the rest of our analysis. Let us now turn to how these measures can be used to describe differences between the pure SM, and the SMEFT. 


\section{Magic for top quarks in the SMEFT}
\label{sec:magic}

In this section, we extend the analysis of ref.~\cite{White:2024nuc}, to present results for the magic of top quark pairs in the SMEFT. Given that our motivation is to compare results with the SM, we will explicitly show results for the difference between the magic in the pure SM and SMEFT cases. Let  $M_2^{(n)}$ denote the contribution to the second Stabliser R\'{e}nyi entropy up to and including terms of ${\cal O}(c^n)$, where $c$ is a generic SMEFT Wilson coefficient. That is, $M_2^{(0)}$ constitutes the SM value for the magic, whereas $M^{(1)}_2$ and $M_2^{(2)}$ constitute the magic including SMEFT-SM interference terms up to linear and quadratic order in the Wilson coefficients respectively.  
We will consider the effect of various SMEFT operators in turn, restricting our attention to operators that contribute already at linear order in the EFT. 

\subsection{Differential results}
\label{sec:differential}

\begin{figure}
    \begin{center}
        \scalebox{0.5}{\includegraphics{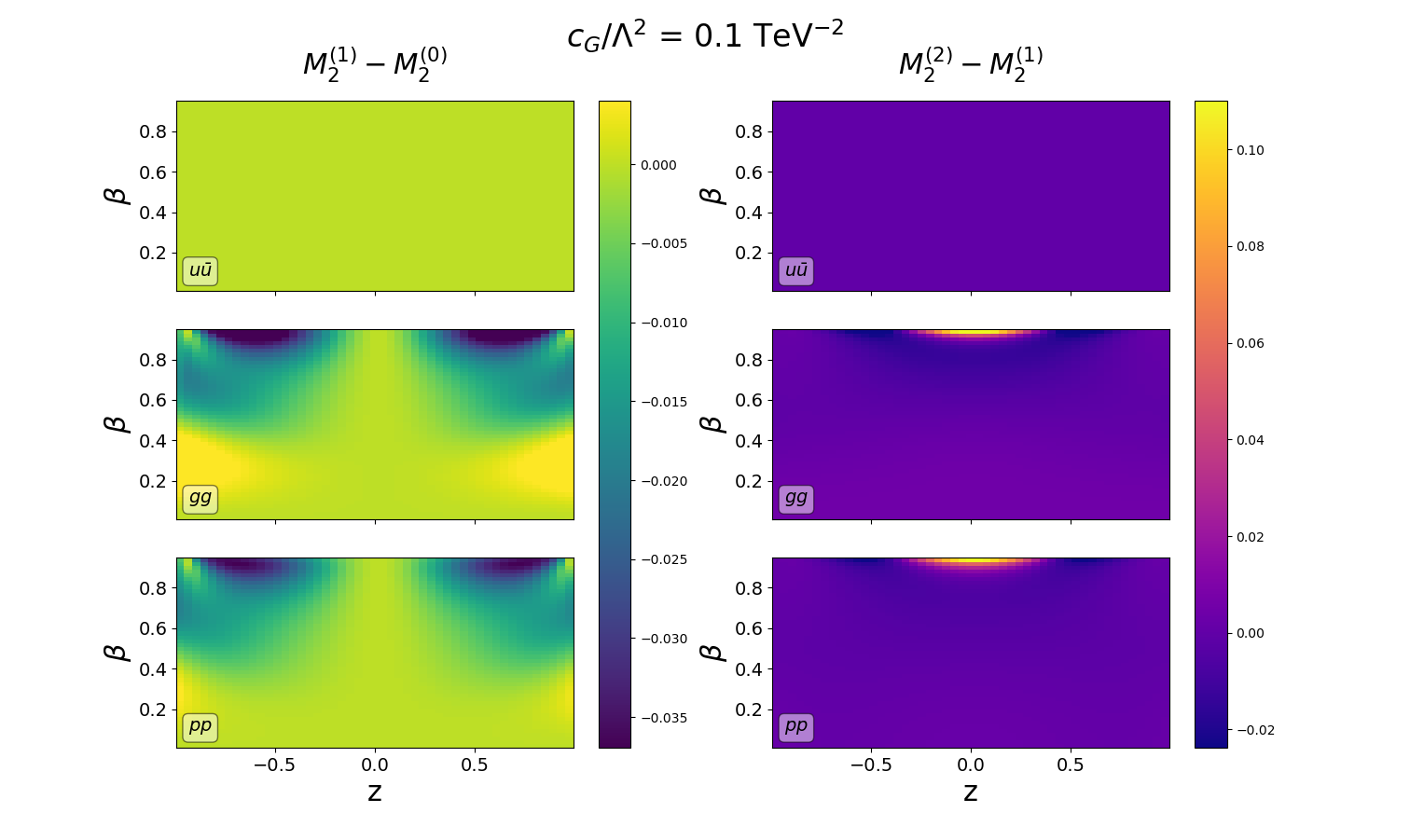}}
        \caption{Left panels: Difference in magic (as measured by the second stabiliser R\'{e}nyi entropy $M_2$) between the SMEFT and SM, for a single non-zero Wilson coefficient $c_G$, and including the contribution that is linear in $c_G$ only. Right panel: Difference between the linearised magic, and the contribution that is quadratic in $c_G$.}
        \label{fig:cG}
    \end{center}
\end{figure}

\begin{figure}
    \begin{center}
        \scalebox{0.5}{\includegraphics{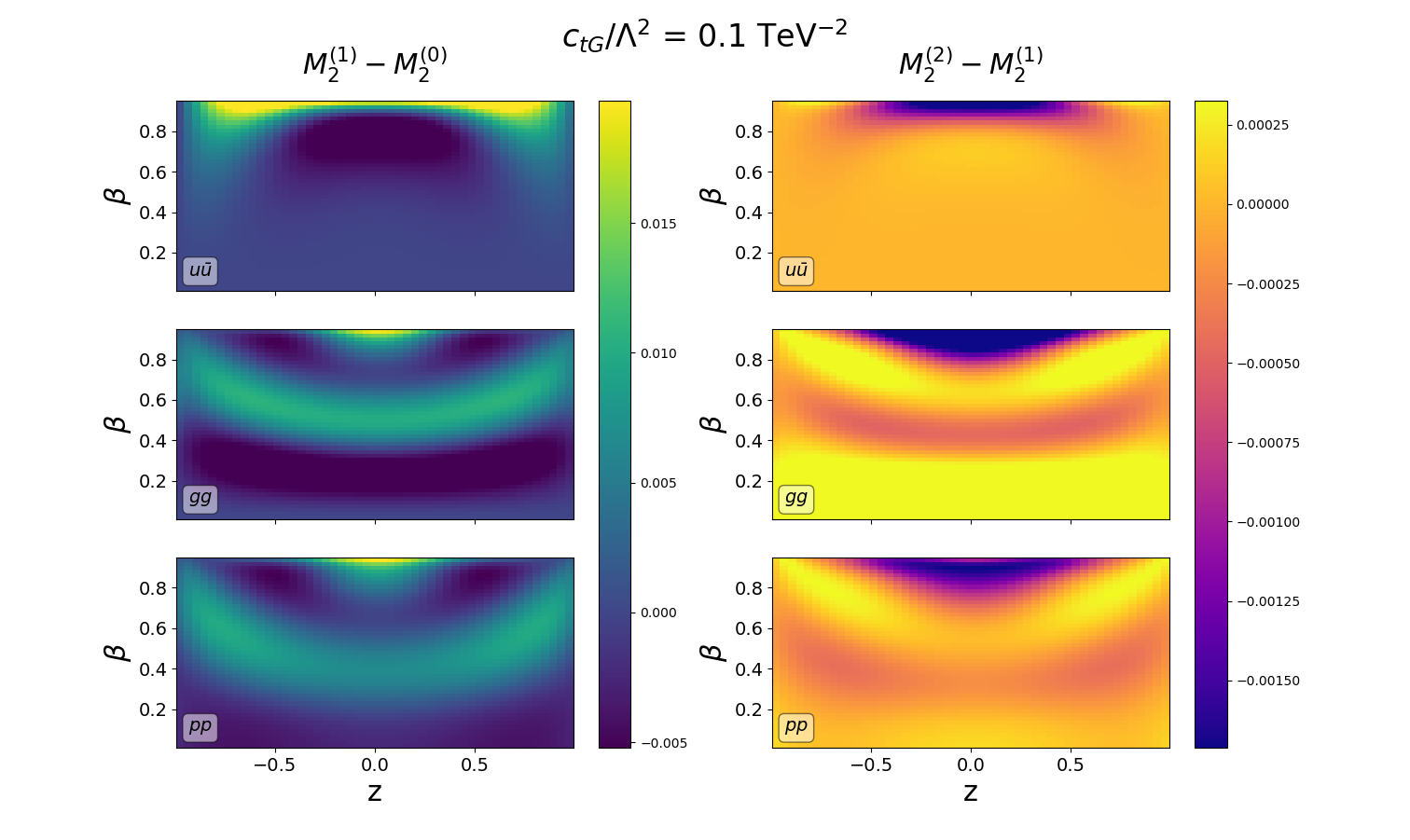}}
        \caption{Left panels: Difference in magic (as measured by the second stabiliser R\'{e}nyi entropy $M_2$) between the SMEFT and SM, for a single non-zero Wilson coefficient $c_{tG}$, and including the contribution that is linear in $c_{tG}$ only. Right panel: Difference between the linearised magic, and the contribution that is quadratic in $c_{tG}$.}
        \label{fig:ctG}
    \end{center}
\end{figure}

\begin{figure}
    \begin{center}
        \scalebox{0.5}{\includegraphics{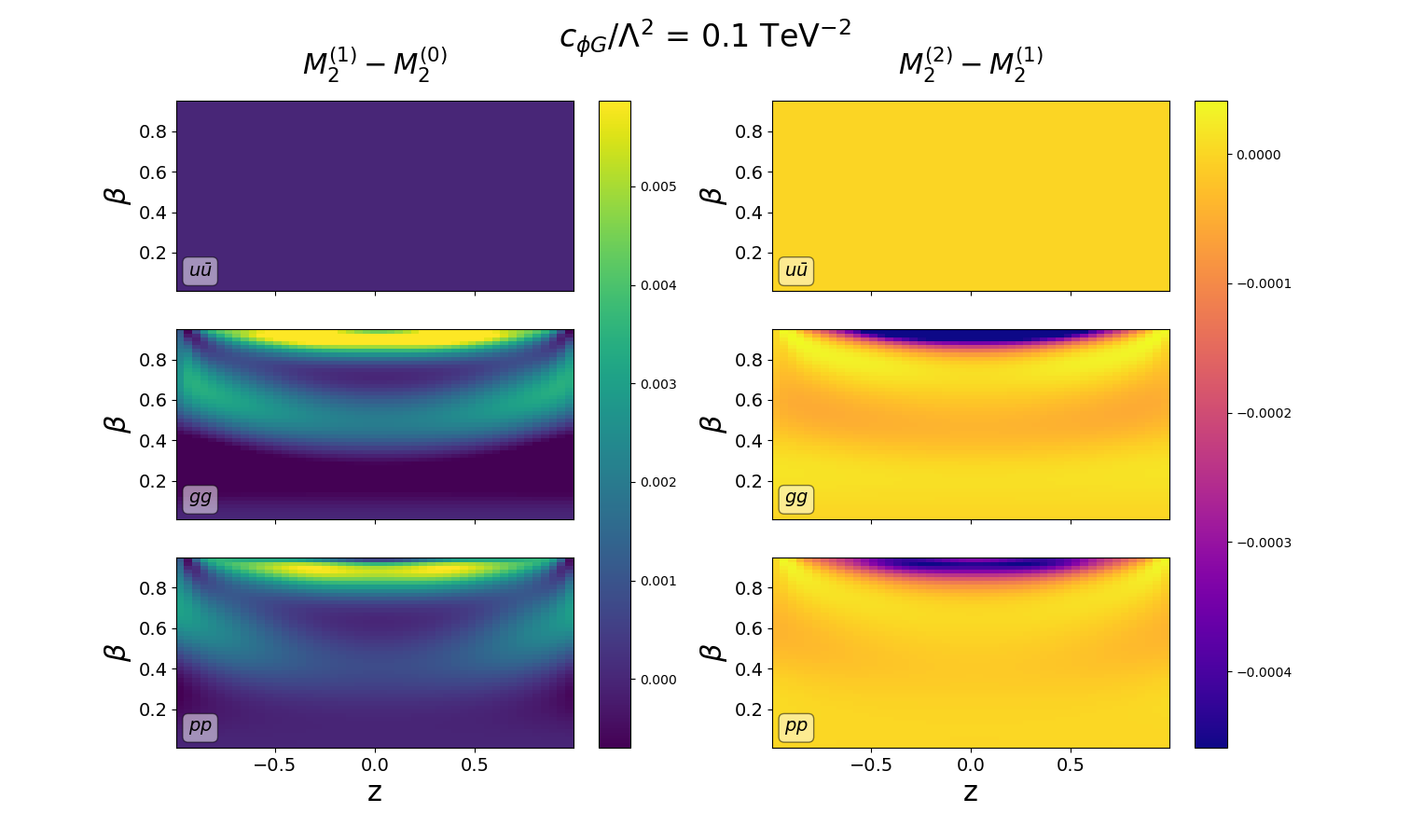}}
        \caption{Left panels: Difference in magic (as measured by the second stabiliser R\'{e}nyi entropy $M_2$) between the SMEFT and SM, for a single non-zero Wilson coefficient $c_{\phi G}$, and including the contribution that is linear in $c_{\phi G}$ only. Right panel: Difference between the linearised magic, and the contribution that is quadratic in $c_{\phi G}$.}
        \label{fig:cphiG}
    \end{center}
\end{figure}

Let us start by considering results fully differentially in $z=\cos\theta$ (where $\theta$ is the scattering angle), and the top quark velocity $\beta$. Results for non-zero Wilson coefficients $\bar{c}_{G}$, $\bar{c}_{tG}$ and $\bar{c}_{\phi G}$ are shown in figures~\ref{fig:cG}, \ref{fig:ctG} and \ref{fig:cphiG} respectively, where we have chosen a value of $0.1 {\rm TeV}^{-2}$ in each case, with all other coefficients set to zero. The left panels in each figure show the difference between the magic including terms up to linear order in the Wilson coefficient, and the pure SM result. In other words, the plots show the contribution to the magic from the linear SMEFT-SM interference term only. The right-hand panels in each figure show the additional contribution that results from terms in the magic that are quadratic in the non-zero Wilson coefficient, which is evident in the much smaller scale. For each figure, the upper row shows results for the $u\bar{u}$ channel only, as a representative of the $q\bar{q}$ initial state. The middle row shows results for the gluon channel, and the lower row shows the full proton level result, including the parton distribution functions. We work throughout in the helicity basis using the expressions for the Fano coefficients given in appendix~\ref{app:F}.

For $c_{G}$ and $c_{\phi G}$, there is no difference in the magic in the $u\bar{u}$ channel, corresponding to the fact that these operators only contribute to the $gg$ channel at linear order. Interestingly, even in channels for which operators do contribute in general, there are kinematic regions in which the magic does not change. This is seen, for example, in the $gg$ channel at threshold ($\beta\rightarrow 0$) for $c_G$ and $c_{\phi G}$, which can be traced to the fact that their corresponding contributions to the Fano coefficients -- reported here in appendix~\ref{app:Fano} -- are weighted by an explicit factor of $\beta^2$, itself originating from the additional derivatives in the field strength tensors in eq.~(\ref{O1}). Moreover, ref.~\cite{Aoude:2022imd} showed that the top quarks remain in the same maximally entangled state as in the pure SM $gg$ channel at threshold, even once the other Wilson coefficients are included. As argued in ref.~\cite{White:2024nuc}, this particular state happens to be a stabiliser state, and thus the SMEFT will not change the fact that the magic vanishes at threshold. 

At quadratic order in the $gg$ channel, the $c_{\phi G}$ contribution vanishes at threshold, but the $\mathcal{O}_{G}$ and $\mathcal{O}_{t G}$ operators induce a triplet state on top of the singlet state produced in the SM. As shown in ref.~\cite{Aoude:2022imd}, one can write the density matrix as 
\begin{equation}
    \rho_{gg}^{\rm SMEFT}(0,z)=p_{gg}|\Psi^+\rangle_\mathbf{p}\langle\Psi^+|_\mathbf{p}+(1-p_{gg})|\Psi^-\rangle_\mathbf{p}\langle\Psi^-|_\mathbf{p}~,
    \label{rhoggSMEFT}
\end{equation}
where $|\Psi^+\rangle_{\vec{p}}$ ($|\Psi^-\rangle_{\vec{p}}$) is a spin triplet (singlet) state along the beam axis $\vec{p}$, and
\begin{equation}
    p_{gg}=\frac{72m_t^2(3\sqrt{2}m_tc_G+vc_{tG})}
    {7\Lambda^4}~,
\end{equation}
where as before $v$ denotes the Higgs vacuum expectation value (VEV). For generic non-zero Wilson coefficients $(c_G, c_{tG})$, eq.~(\ref{rhoggSMEFT}) constitutes a mixed state rather than the pure state obtained in the pure SM, and thus the magic increases. The largest changes in the magic in this $gg$ channel, however, are seen at high $\beta$. Again this can be understood from the fact that a pure maximally entangled state in the SM (at least for the central region $\theta=\pi/2$) gets replaced by a mixed state once SMEFT effects are included. One also expects effective operators (which can contain additional derivatives giving rise to extra powers of velocity or energy in momentum space) to be more significant at higher $\beta$. Care, however, is needed in interpreting results at high $\beta$, given that the SMEFT is only valid for partonic centre of mass energies which are less than the cut-off scale i.e.
\begin{equation}
    \hat{s}<\Lambda^2~.
    \label{shatbound}
\end{equation}
Assuming a unit Wilson coefficient and a top quark mass of $m_t=173$ GeV, this bound amounts to $\beta<0.962$ in figures~\ref{fig:cG}--\ref{fig:cphiG}. We then see that, over the remaining phase space, quadratic SMEFT corrections to the magic are small, which itself implies that the EFT expansion is behaving well.

Having discussed the $gg$ channel, let us now turn to the 
$q\bar{q}$ channels. These receive contributions from $c_{tG}$, and from the particular combinations of four-fermion operators outlined in appendix~\ref{app:Fano}. In figure~\ref{fig:ctG}, we see a correction to the magic from $c_{tG}$ is indeed observed, but that this is zero at threshold ($\beta\rightarrow 0$). This can be explained from the fact that SMEFT contributions lead to the mixed separable state
\begin{equation}
    \rho_{q\bar{q}}^{\rm SMEFT}=p_{q\bar{q}}
    |\uparrow\uparrow\rangle_{\vec{p}}\langle \uparrow\uparrow|_{\vec{p}}+(1-p_{q\bar{q}})
    |\downarrow\downarrow\rangle_{\vec{p}}\langle \downarrow\downarrow|_{\vec{p}}~,
    \label{qqstate}
\end{equation}
near threshold~\cite{Aoude:2022imd}, where spin states are again defined relative to the beam direction, and 
\begin{equation}
    p_{q\bar{q}}=\frac12-\frac{4c_{VA}^{(8),u}}{\Lambda^2}+{\cal O}(\Lambda^{-4})~. 
    \label{pqqbar}
\end{equation}
The coefficient appearing on the right-hand side is defined in eq.~(\ref{cVV}), and consists of a linear combination of four-fermion operators with no contribution from $c_{tG}$. Furthermore, if $p_{q\bar{q}}=1/2$, then the state of eq.~(\ref{qqstate}) turns out to be a stabiliser state~\cite{White:2024nuc}, whose magic necessarily vanishes. Corrections to the magic from $c_{tG}$ are moderate over the remaining parameter space in the quark channel, where the more pronounced difference at very high $\beta$ falls in the region where the EFT may not be valid. 

\begin{figure}
    \begin{center}
        \scalebox{0.5}{\includegraphics{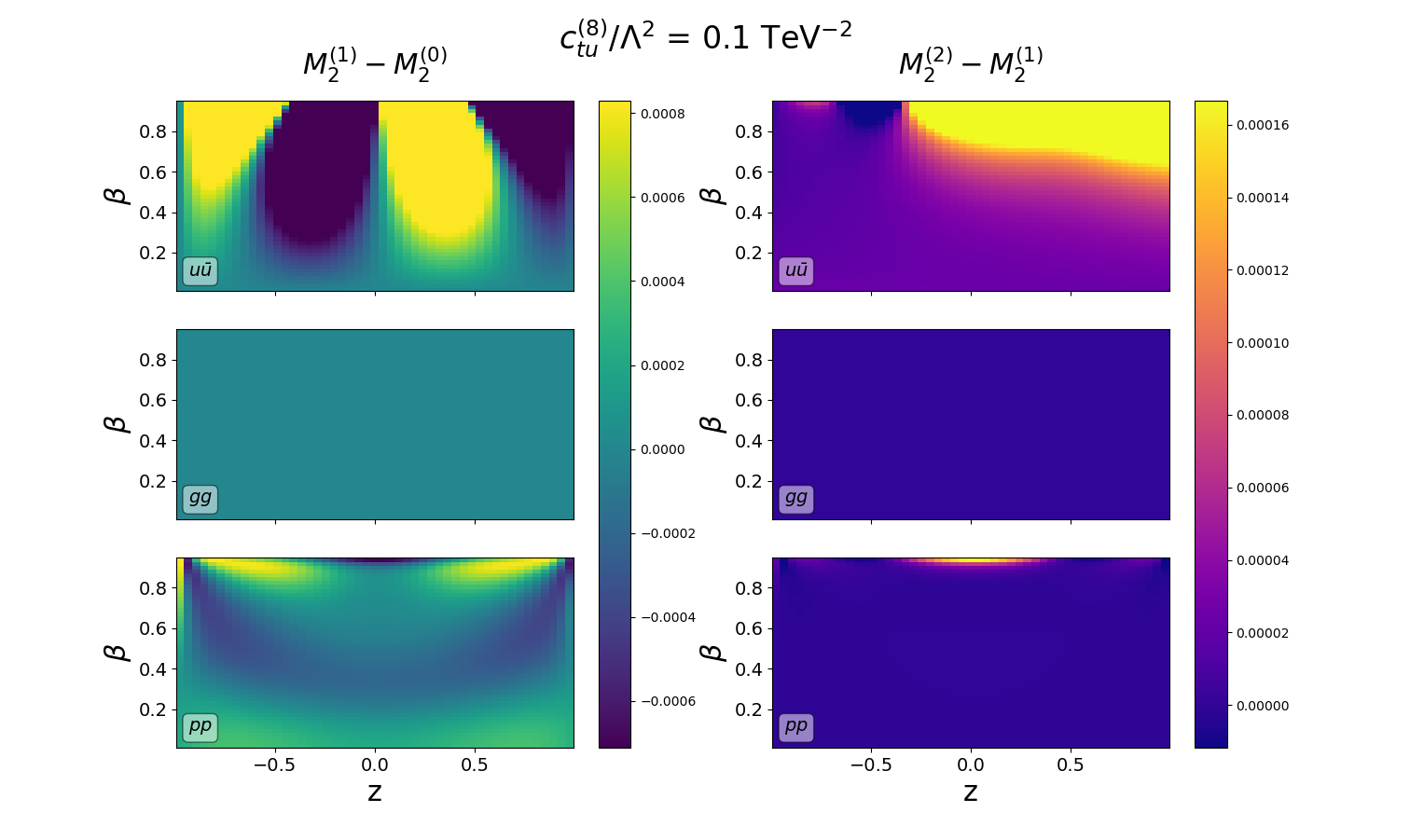}}
        \caption{Left panels: Difference in magic (as measured by the second stabiliser R\'{e}nyi entropy $M_2$) between the SMEFT and SM, for a single non-zero Wilson coefficient $c_{tu}^{(8)}$, and including the contribution that is linear in $c_{tu}^{(8)}$ only. Right panel: Difference between the linearised magic, and the contribution that is quadratic in $c_{tu}^{(8)}$.}
        \label{fig:ctu8}
    \end{center}
\end{figure}

As a representative example of the effect of four-fermion operators, figure~\ref{fig:ctu8} shows the correction to the magic resulting from a non-zero $c_{tu}^{(8)}$ (with all other Wilson coefficients zero). There is no contribution to the $gg$ channel as expected. For the $u\bar{u}$ channel, the correction to the magic is low near threshold, as expected from the above remarks: the final state becomes a stabiliser state at threshold in the SM, and the Wilson coefficient proves to be a small correction to this. Away from threshold -- and well below the upper bound of $\beta\simeq 0.96$ at which the EFT validity should be questioned -- there is a highly interesting pattern, in which corrections to the magic can be positive or negative depending on the scattering angle.  The differing signs of corrections to the magic opens up the possibility of cancellations if different operator contributions are combined. As for the $gg$ channel, we find in both figures~\ref{fig:ctG} and~\ref{fig:ctu8} that the quadratic corrections are negligible compared to the linear corrections over most of the phase space, which suggests that higher-order EFT corrections are under control. 

As well as individual partonic channels, we have also shown results for proton-proton initial states in figures~\ref{fig:cG}--\ref{fig:ctu8}, by combining with the appropriate parton distribution functions (PDFs) as per equation~\ref{Cpp}, similar to the pure SM results of ref.~\cite{White:2024nuc}. As for the results of that reference, one typically sees that the proton-level results are strongly determined by the gluon channel, as expected given the dominance of the gluon PDF at the LHC. This effect is arguably most pronounced for the four-fermion operators, where e.g. the striking pattern of magic corrections in the upper-left panel of figure~\ref{fig:ctu8} is significantly washed out in the lower-left panel.

\subsection{Angular-averaged results}
\label{sec:average}

\begin{figure}
        \centering
        \begin{subfigure}[b]{0.475\textwidth}
            \centering
            \includegraphics[width=\textwidth]{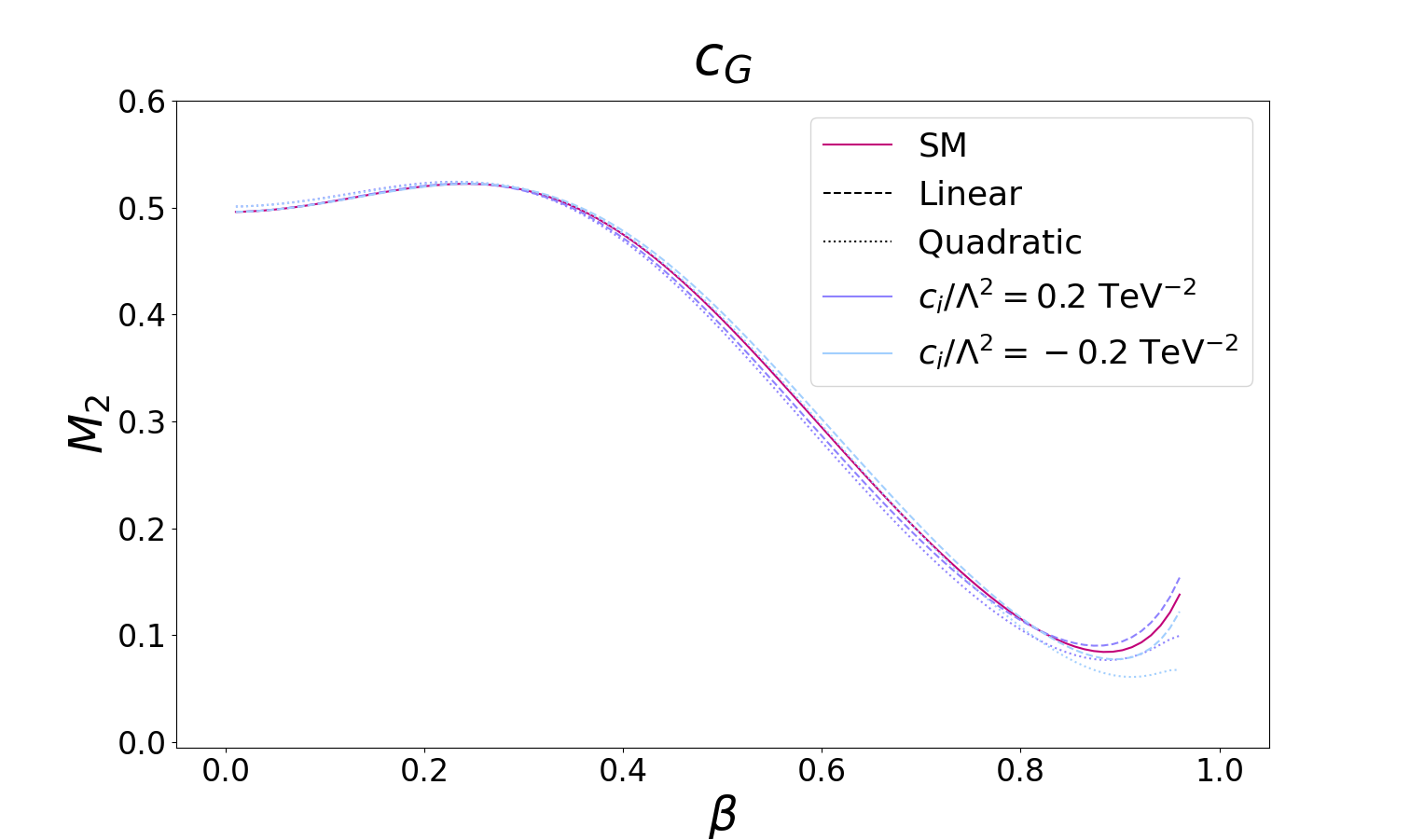}
            \label{fig:cG_av}
        \end{subfigure}
        \hfill
        \begin{subfigure}[b]{0.475\textwidth}  
            \centering 
            \includegraphics[width=\textwidth]{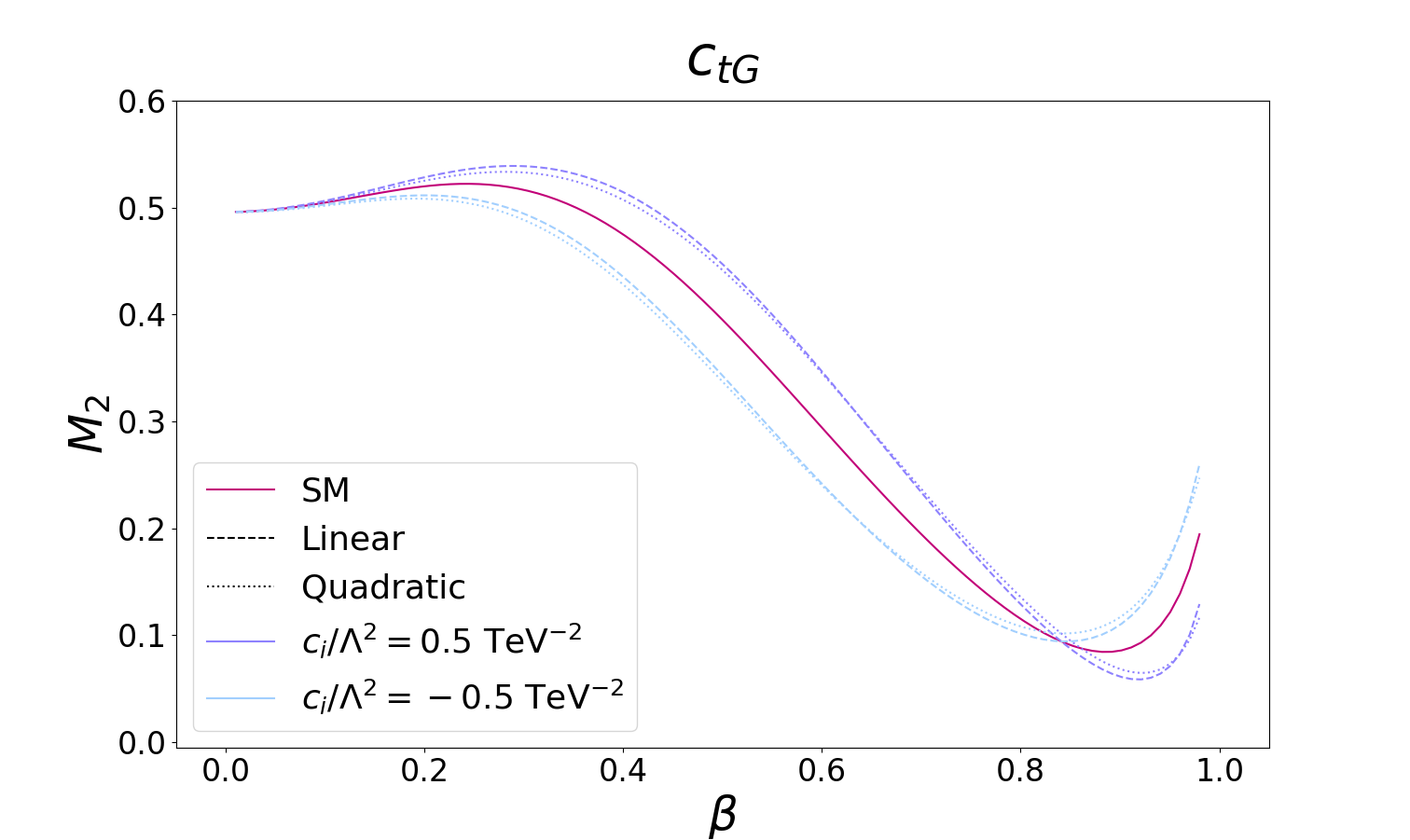}
            \label{fig:ctG_av}
        \end{subfigure}
        \vskip\baselineskip
        \begin{subfigure}[b]{0.475\textwidth}   
            \centering 
            \includegraphics[width=\textwidth]{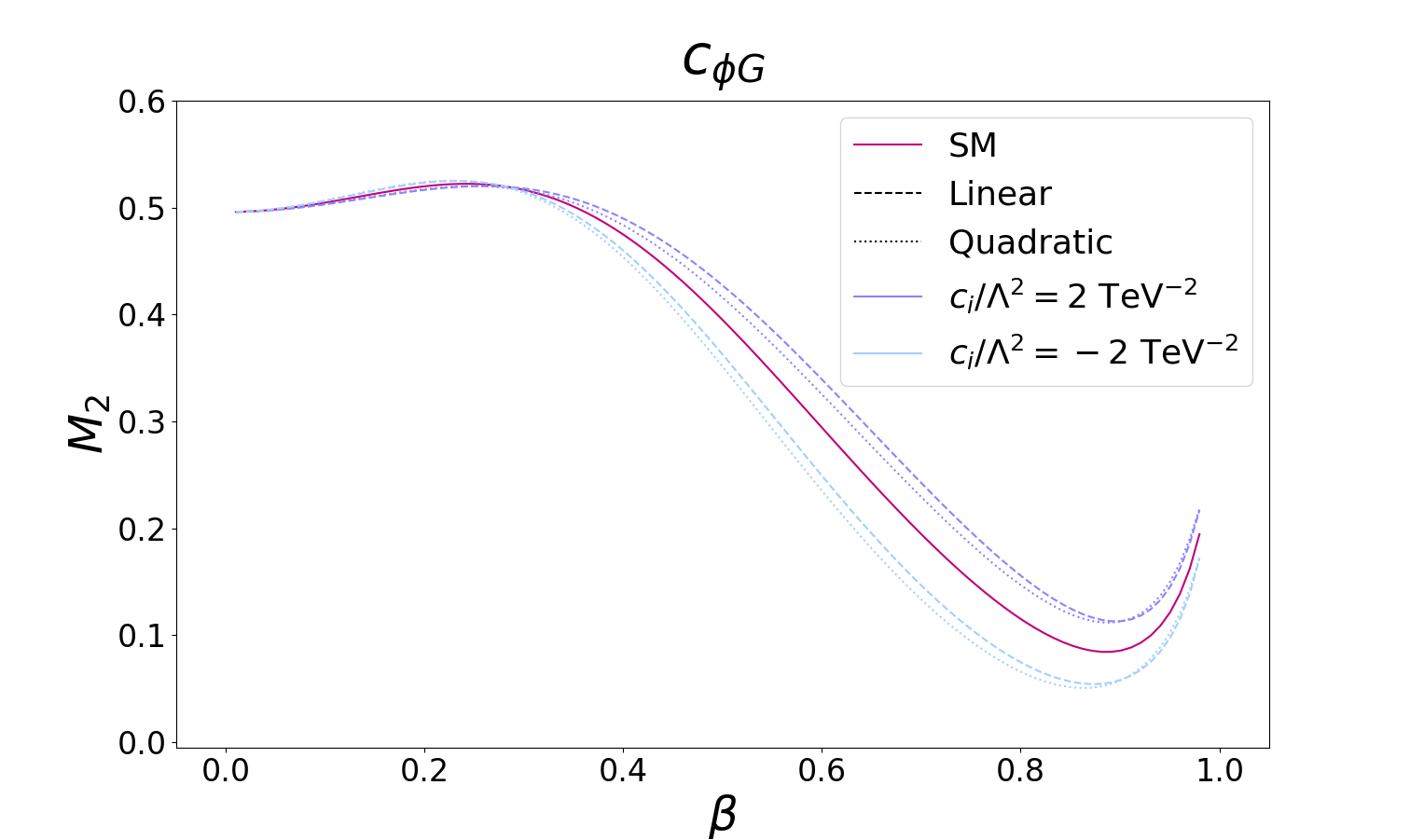}
            \label{fig:cpG_av}
        \end{subfigure}
        \hfill
        \begin{subfigure}[b]{0.475\textwidth}   
            \centering 
            \includegraphics[width=\textwidth]{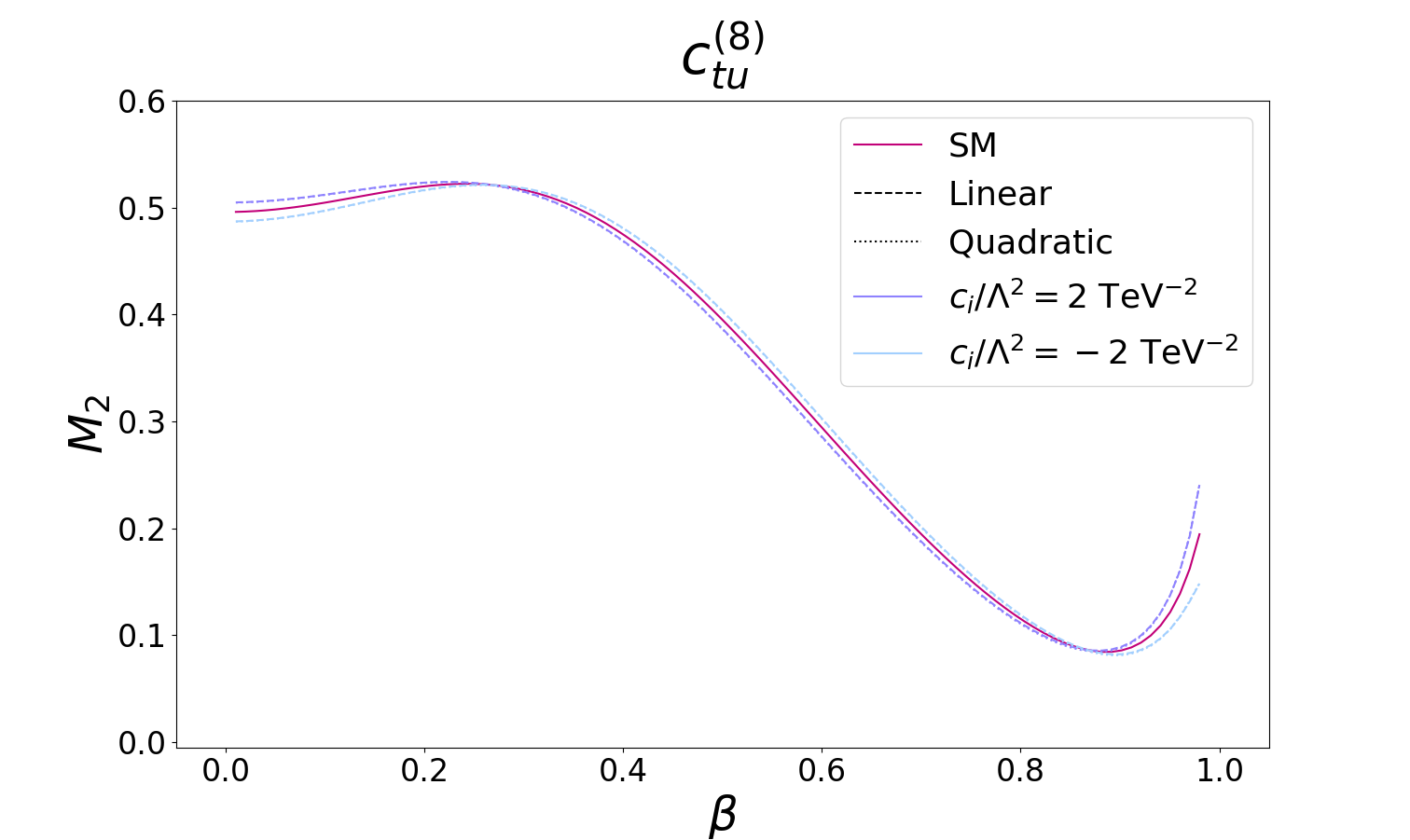}
            \label{fig:ctu8_av}
        \end{subfigure}
        \caption
        {Angular-averaged result for the magic in the beam basis, for non-zero SMEFT Wilson coefficients as shown, and all other coefficients set to zero. } 
        \label{fig:angularmagic}
\end{figure}

As well as fully differential results, it is also instructive to see how the magic behaves when averaged over all scattering angles (N.B. similar analyses have been performed for entanglement measures and SM magic in refs.~\cite{Aoude:2022imd,White:2024nuc}). In SM processes, it has been observed that angular averaging generically increases the magic, largely due to the fact that more mixed states are obtained, taking one further away from particular stabiliser states. In figure~\ref{fig:angularmagic}, we compare the SMEFT results with the SM for the same four operators as in figs.~\ref{fig:cG}--\ref{fig:ctu8}, using higher absolute values for the Wilson coefficients where necessary in order to make differences show up more clearly.   We do not show results for colour-singlet four-fermion operators, which do not contribute at dimension six in the SMEFT expansion. Quadratic insertions of these operators also lead to very small corrections, due to the fact that the $q\bar{q}$ channel is subdominant to the $gg$ channel at the LHC. Our choice of the four-fermion Wilson coefficient $c_{tu}^{(8)}$ is representative of results for other coefficients, given that only the particular linear combinations of eq.~(\ref{cVV}) arise in top pair production. As in ref.~\cite{White:2024nuc}, the relevant angular-averaged coefficients are calculated in the fixed beam basis, rather than the helicity basis, so that one may refer spins to spatial directions that survive after averaging over all possible top quark directions.  Expressions for the angularly averaged Fano coefficients in this basis are provided in appendix~\ref{app:ang}. In each plot, we compare the SM result with results obtained from positive and negative Wilson coefficients with the same magnitude. We further show the effect of linearising the SMEFT contributions, vs. expanding the magic to quadratic order in the SMEFT coefficient as detailed above.

In figure~\ref{fig:angularmagic}, we notice that the effect of our example four-fermion operator is very small, both at linear and quadratic order. This is commensurate with our above statement that at LHC energies one is dominated by the $gg$ channel, which tends to downweight effects arising from the $q\bar{q}$ initial state. One may also compare this with fig.~\ref{fig:ctu8}, in which a significant profile of corrections to the magic for the $q\bar{q}$ channel is washed out in the proton-level result. Unsurprisingly then, differences in fig.~\ref{fig:angularmagic} are more pronounced for the set of operators $\{c_G, c_{tG}, c_{\phi G}\}$ that contribute in the $gg$ channel. We see that corrections tend to zero at threshold, consistent with the remarks of the previous section: whilst the results in figs.~\ref{fig:cG}--\ref{fig:cphiG} are in a different basis, it nevertheless remains true that the top pairs are in a stabiliser state at threshold to a good approximation. For $c_{\phi G}$, we see that corrections also vanish in the high-velocity limit $\beta\rightarrow 1$, which can be traced to the analytic behaviour of the angular-averaged coefficients in this limit~\cite{Aoude:2022imd}. For moderate $\beta$ values, we note that the largest effects originate from $c_{tG}$, matching what was seen for fully differential results (albeit in a different basis) in the previous section. Overall, quadratic results are a small correction to the linear results, suggesting that higher orders in the EFT expansion are under control. 

The changes to magic induced by SMEFT operators imply that the magic can indeed be used to probe new physics effects, and we will return to this question below. Before doing so, however, let us examine the other QI observables defined in section~\ref{sec:tracedistance}, which were argued in ref.~\cite{Fabbrichesi:2025ywl} to also be useful in this regard.

\section{Comparison of different QI measures}
\label{sec:QI}

\begin{figure}
        \centering
        \begin{subfigure}[b]{0.475\textwidth}
            \centering
            \includegraphics[width=\textwidth]{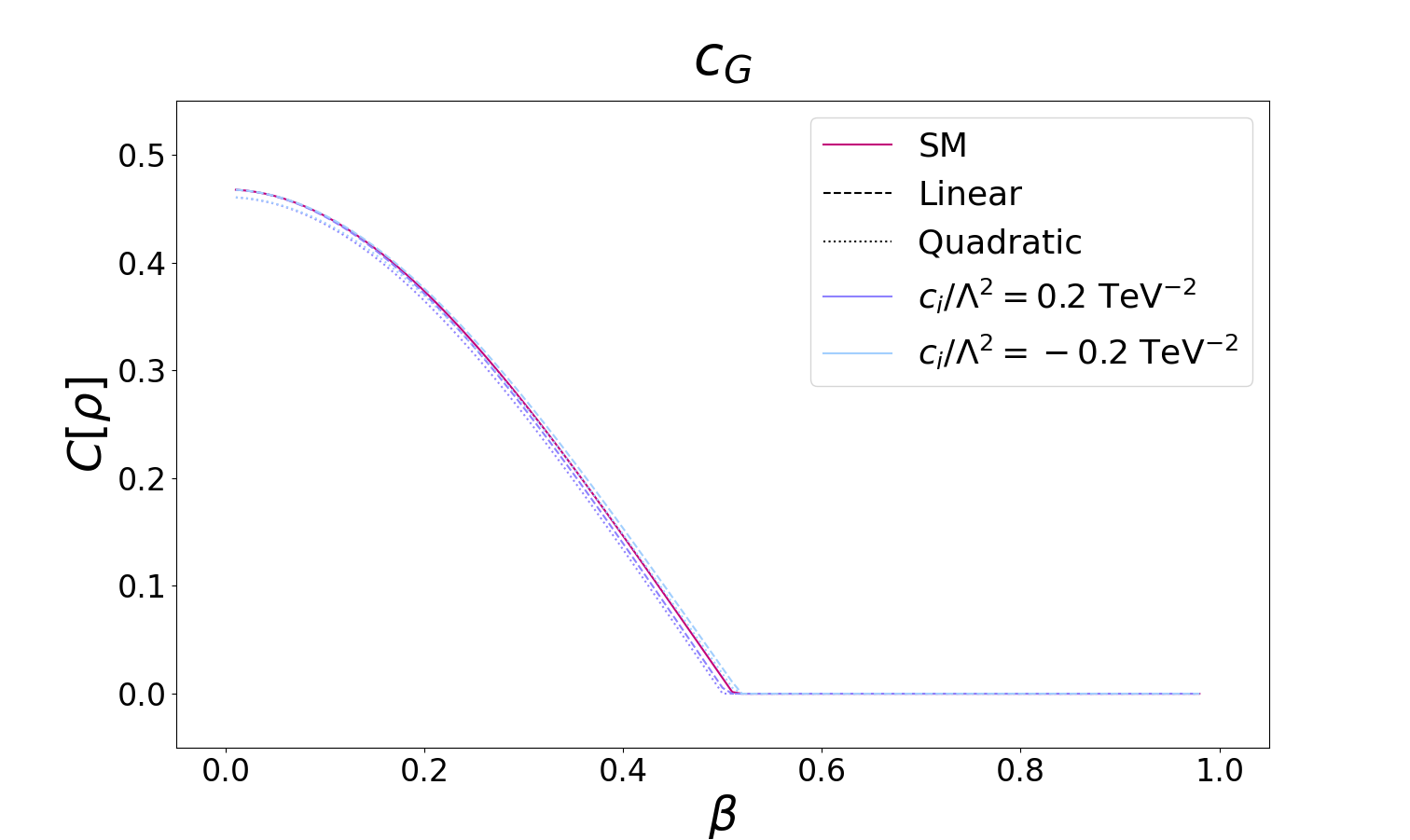}
            \label{fig:concurrence_cG}
        \end{subfigure}
        \hfill
        \begin{subfigure}[b]{0.475\textwidth}  
            \centering 
            \includegraphics[width=\textwidth]{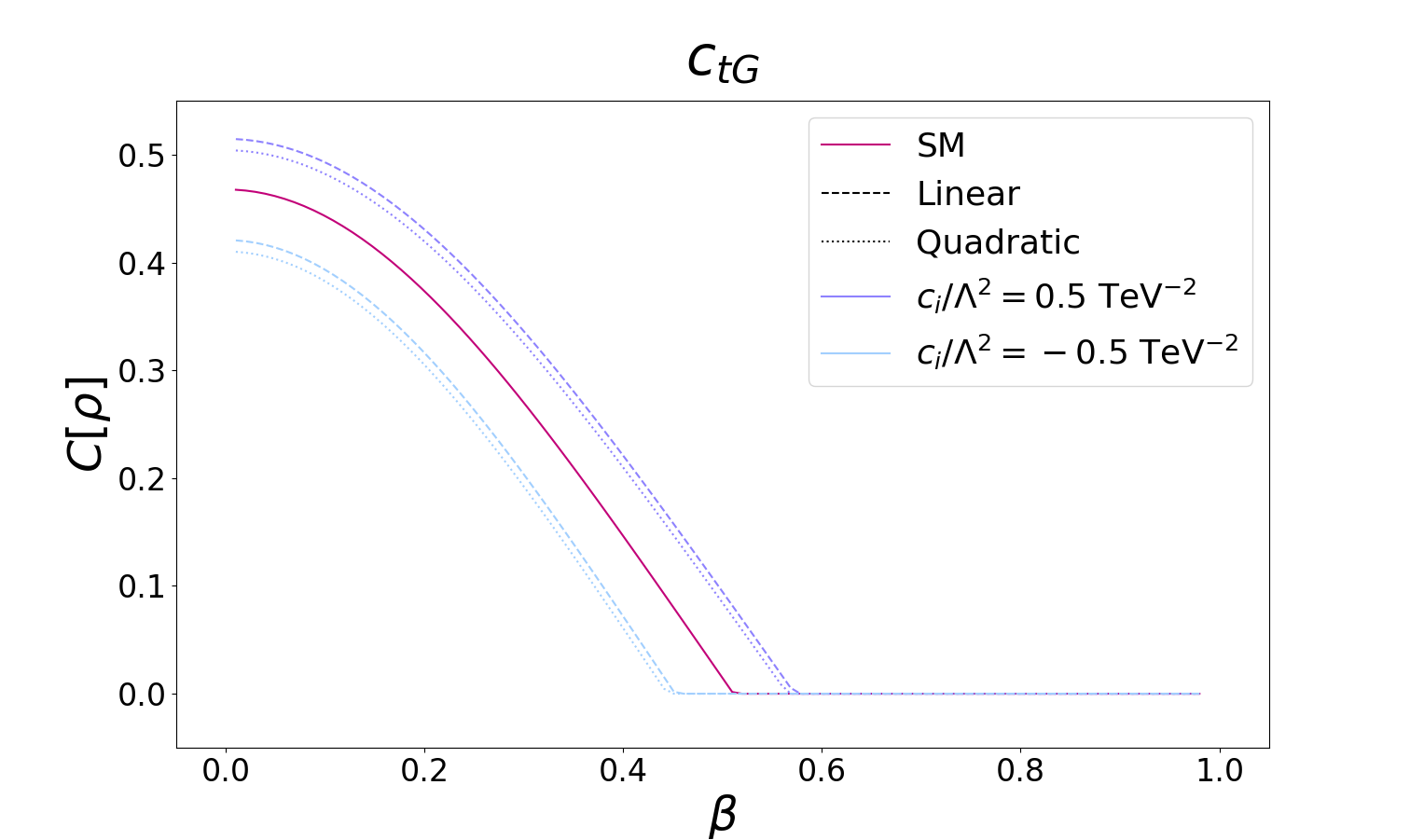}
            \label{fig:concurrence_ctG}
        \end{subfigure}
        \vskip\baselineskip
        \begin{subfigure}[b]{0.475\textwidth}   
            \centering 
            \includegraphics[width=\textwidth]{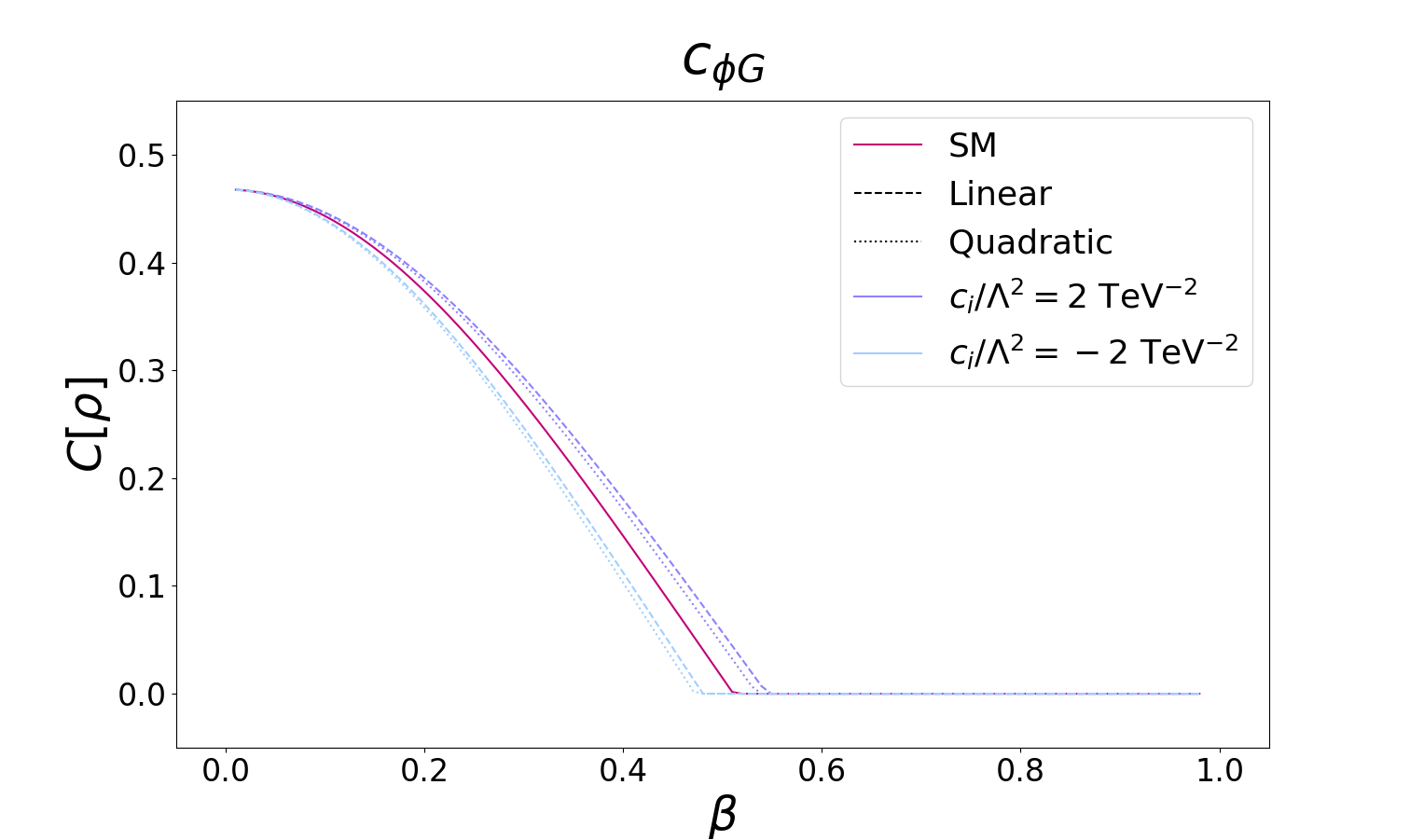}
            \label{fig:concurrence_cpG}
        \end{subfigure}
        \hfill
        \begin{subfigure}[b]{0.475\textwidth}   
            \centering 
            \includegraphics[width=\textwidth]{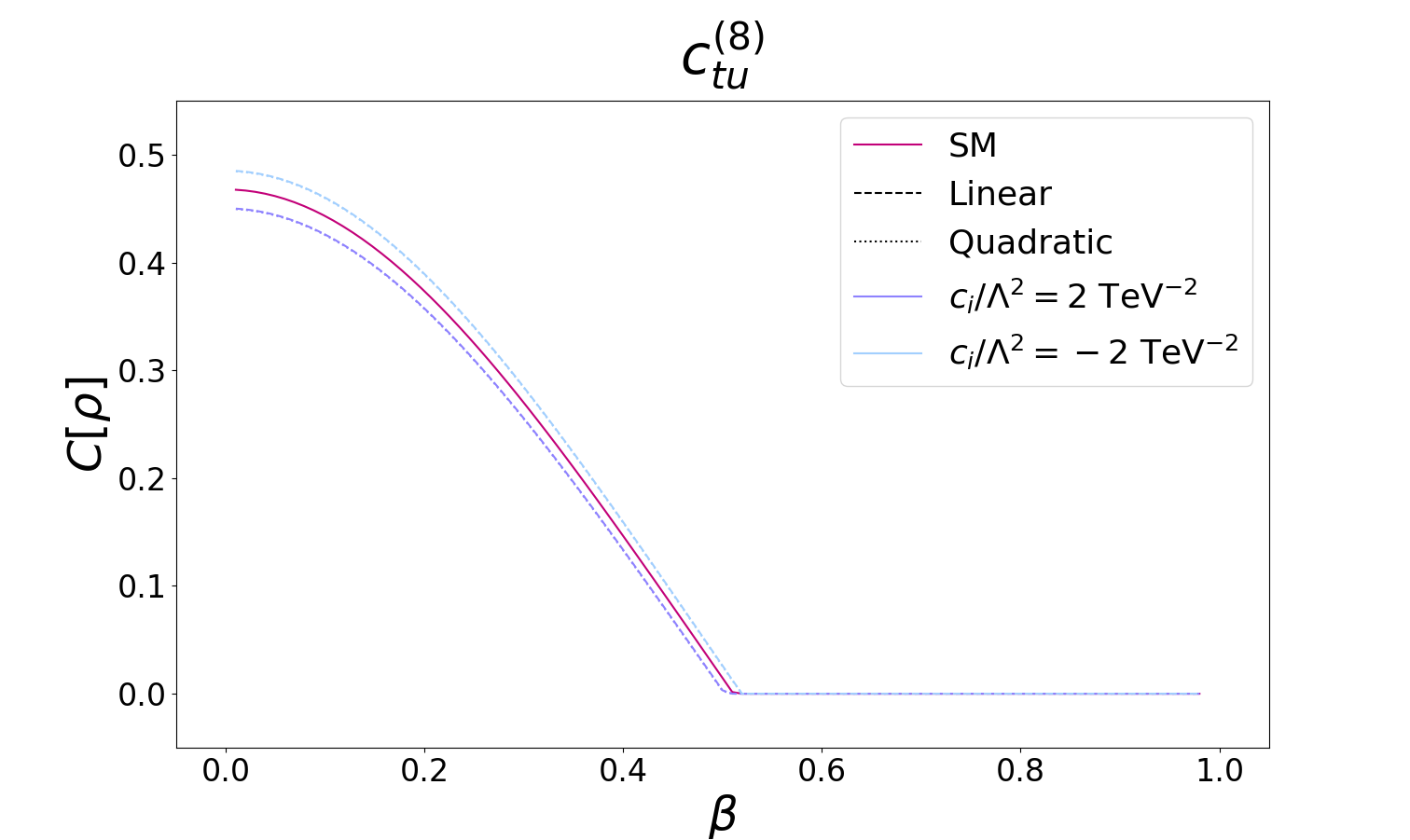}
            \label{fig:concurrence_ctu8}
        \end{subfigure}
        \caption
        {Angular-averaged result for the concurrence in the beam basis, for non-zero SMEFT Wilson coefficients as shown, and all other coefficients set to zero.} 
        \label{fig:concurrence}
\end{figure}

Having examined how the magic behaves, we can also look at the other quantum information measures discussed in section~\ref{sec:tracedistance}. As for the magic, we expand these quantities in the Wilson coefficients, keeping terms up to linear or quadratic order respectively. We begin with the concurrence. Although this has been previously discussed in ref.~\cite{Aoude:2022imd}, we here show a more complete set of angular-averaged results in the beam basis, allowing a straightforward comparison with other measures. The concurrence is shown in figure~\ref{fig:concurrence}, for the same choice of Wilson coefficients as in the magic plot of figure~\ref{fig:angularmagic}. As for the latter plots, we have chosen higher absolute values of each normalised Wilson coefficient where needed, to accentuate differences. For all coefficients, the concurrence tends to zero at high $\beta$ values (i.e. the region of high top pair invariant mass), corresponding to the known fact that there is no entanglement in that region~\cite{Aoude:2022imd}. We see that corrections are small but non-negligible, and that quadratic corrections in the EFT expansion are typically small over the whole parameter space, indicating that the EFT expansion is under control.

\begin{figure}
        \centering
        \begin{subfigure}[b]{0.475\textwidth}
            \centering
            \includegraphics[width=\textwidth]{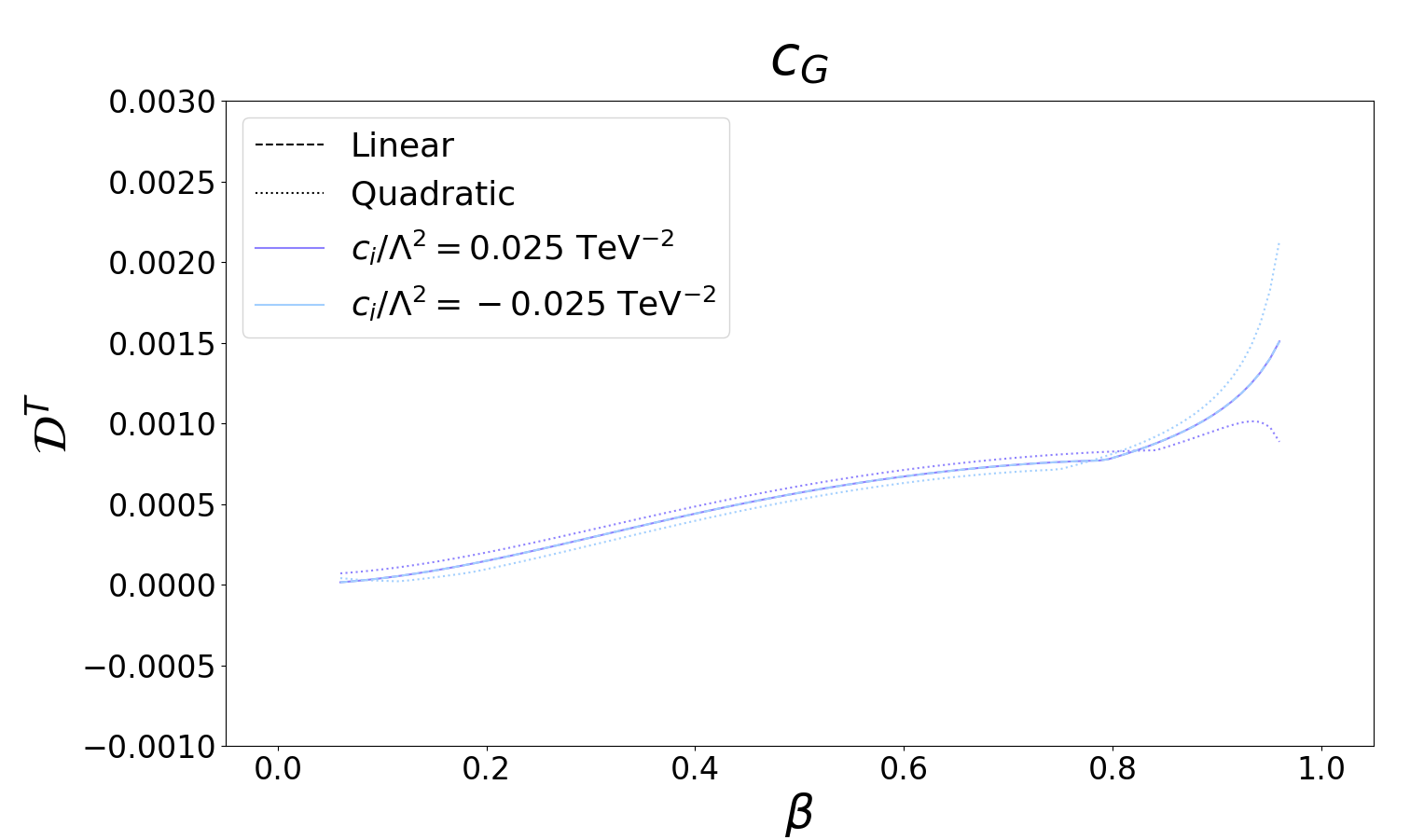}
            \label{fig:td_cG_av}
        \end{subfigure}
        \hfill
        \begin{subfigure}[b]{0.475\textwidth}  
            \centering 
            \includegraphics[width=\textwidth]{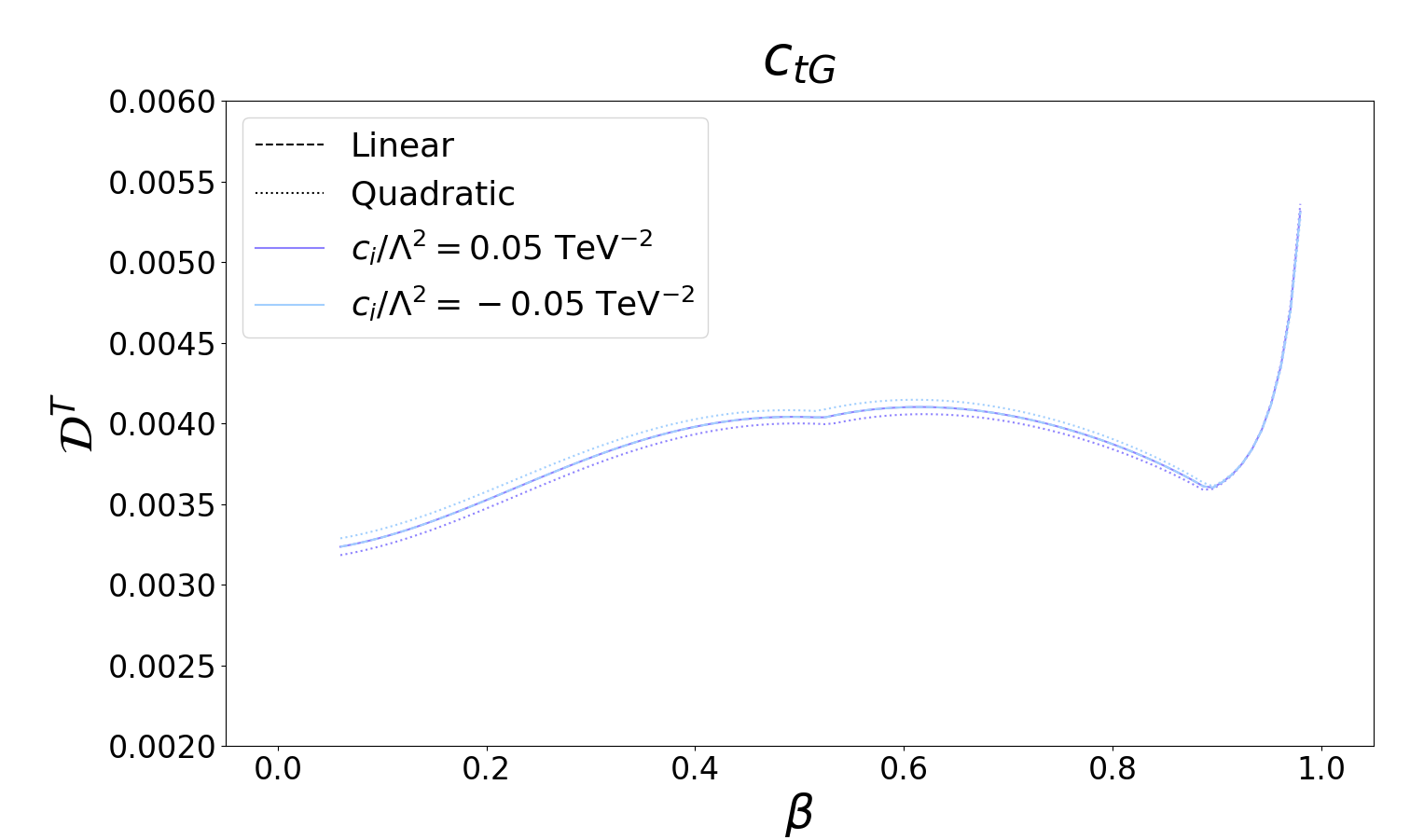}
            \label{fig:td_ctG_av}
        \end{subfigure}
        \vskip\baselineskip
        \begin{subfigure}[b]{0.475\textwidth}   
            \centering 
            \includegraphics[width=\textwidth]{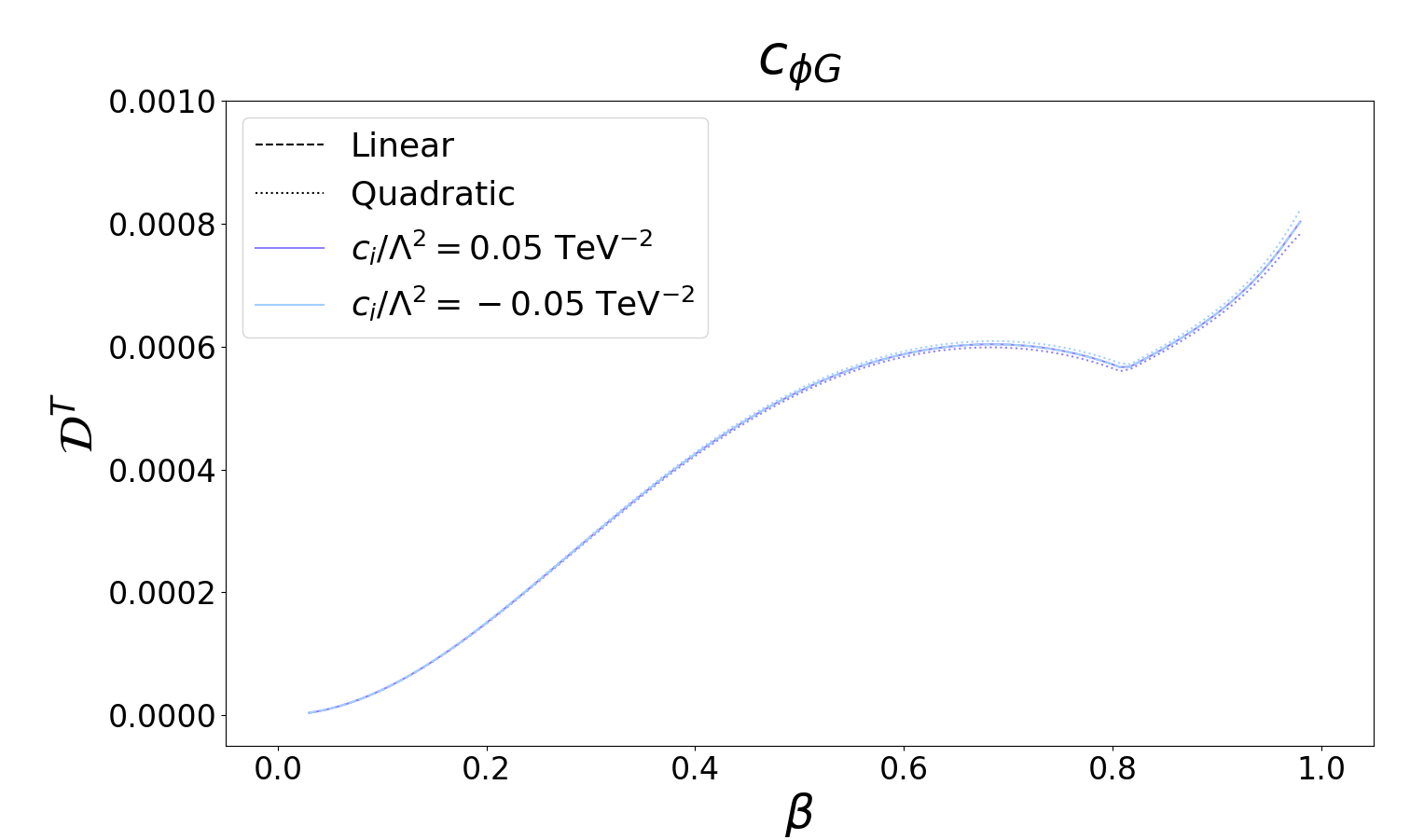}
            \label{fig:td_cpG_av}
        \end{subfigure}
        \hfill
        \begin{subfigure}[b]{0.475\textwidth}   
            \centering 
            \includegraphics[width=\textwidth]{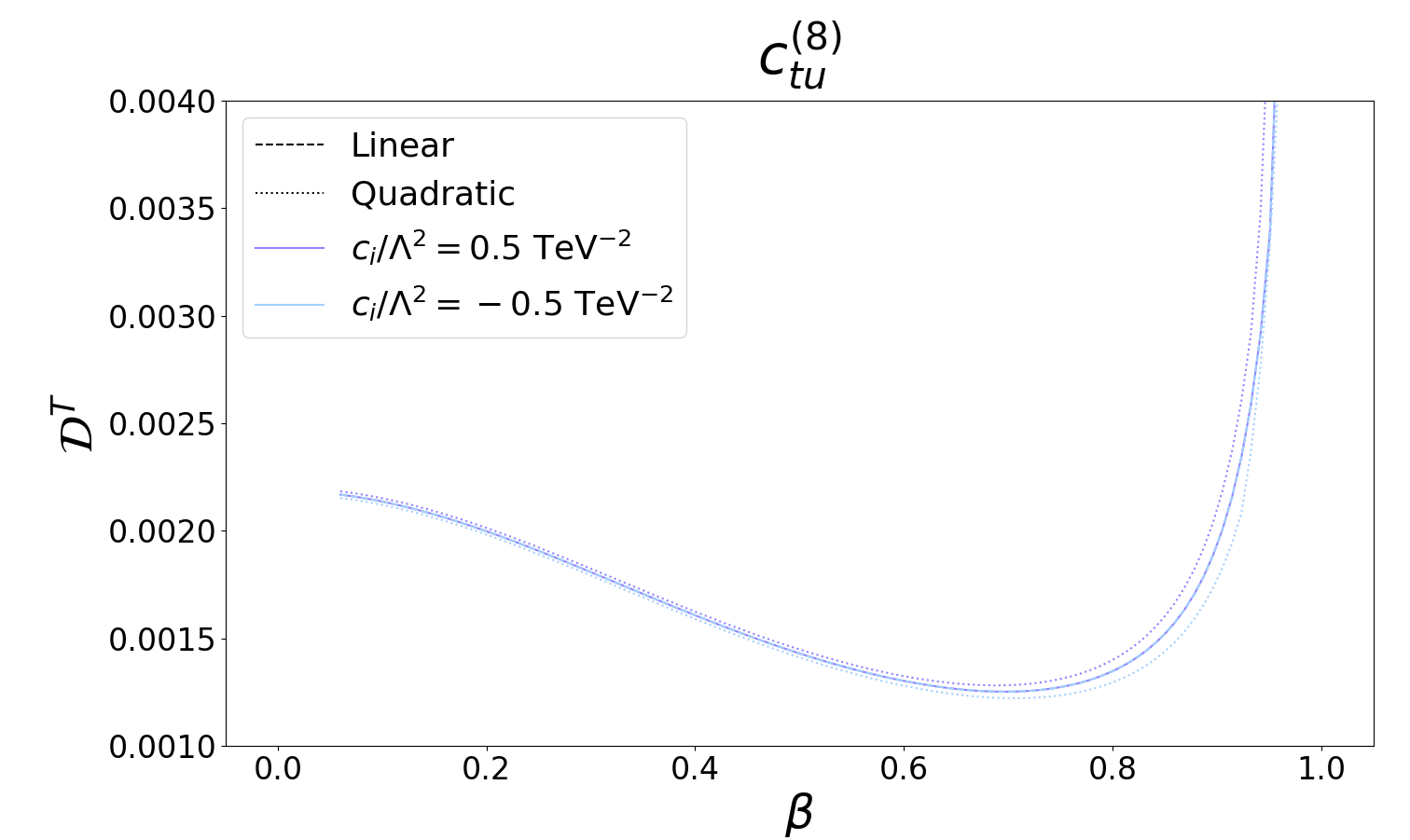}
            \label{fig:td_ctu8_av}
        \end{subfigure}
        \caption
        {Angular-averaged result for the trace distance in the beam basis, for non-zero SMEFT Wilson coefficients as shown, and all other coefficients set to zero.} 
        \label{fig:td}
\end{figure}

As discussed in section~\ref{sec:tracedistance}, both the trace distance and fidelity are explicitly designed to measure the `closeness' of two particular density matrices, which for our purposes can be taken to be the pure SM top quark spin density matrix, and the analagous result including SMEFT corrections. Figure~\ref{fig:td} shows angularly-averaged results for the trace distance (in the beam basis), where we use smaller Wilson coefficients than the previous plots, given that the zero nature of the observable in the SM means that we do not have to enhance the effect for ease of visualisation. For the representative four-fermion coefficient $c_{tu}^{(8)}$, we see that the result is small over most of the parameter space, which is perhaps unsurprising given the similarities with the SM magic observed in figure~\ref{fig:angularmagic}: dominance of the $gg$-channel implies that the final state is approximately SM-like if a single Wilson coefficient is turned on that only contributes in the $q\bar{q}$ channel. At high $\beta$-values, the trace distance diverges due to the analytic behaviour of the Fano coefficients. However, this divergence is unphysical given that it corresponds to the breakdown of the EFT. One can similarly ignore divergences in the other plots in figure~\ref{fig:td}.

Commensurate with the above remarks, we see that the trace distance is in general larger for those operators that contribute in the $gg$ channel. For $c_G$ and $c_{\phi G}$, the trace distance vanishes at threshold ($\beta\rightarrow 0$), and at linear order in the SMEFT, owing to the fact that the top quarks remain in a SM-like stabiliser state. For $c_{tG}$, the trace distance is non-zero but small. This quantifies our earlier observation made for the magic in this region, that it is close to the SM value, owing to the final state being `approximately stabiliser'. By and large, quadratic corrections to the linearised results are small, but in general more pronounced than for the magic for the same values of the Wilson coefficients. This is itself to be expected, given that both the trace distance and fidelity distance are -- by definition -- zero in the SM itself. Thus, the first non-zero contribution to these observables happens at linear order in the SMEFT, such that the quadratic corrections can be proportionally more pronounced. 

\begin{figure}
        \centering
        \begin{subfigure}[b]{0.475\textwidth}
            \centering
            \includegraphics[width=\textwidth]{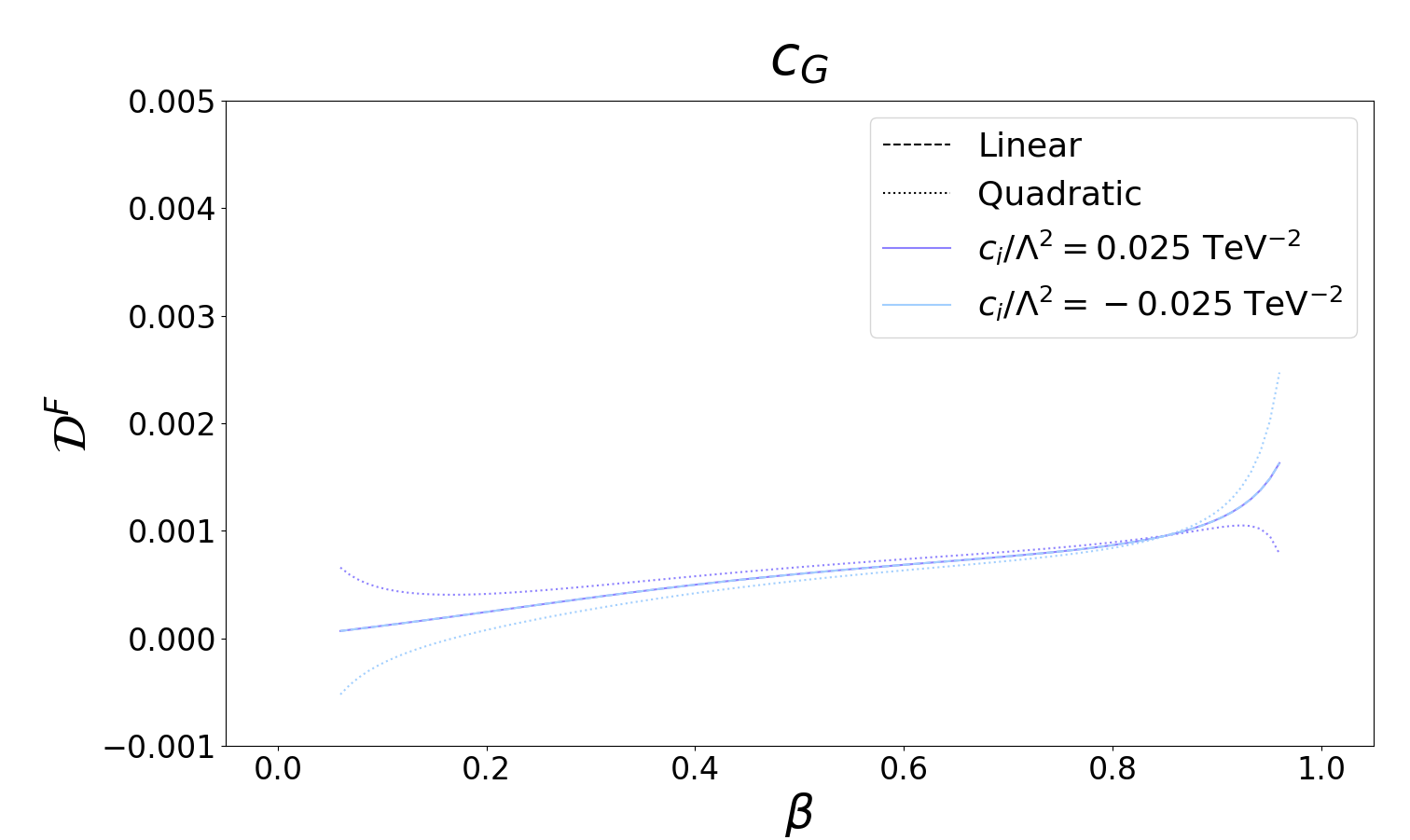}
            \label{fig:fd_cG_av}
        \end{subfigure}
        \hfill
        \begin{subfigure}[b]{0.475\textwidth}  
            \centering 
            \includegraphics[width=\textwidth]{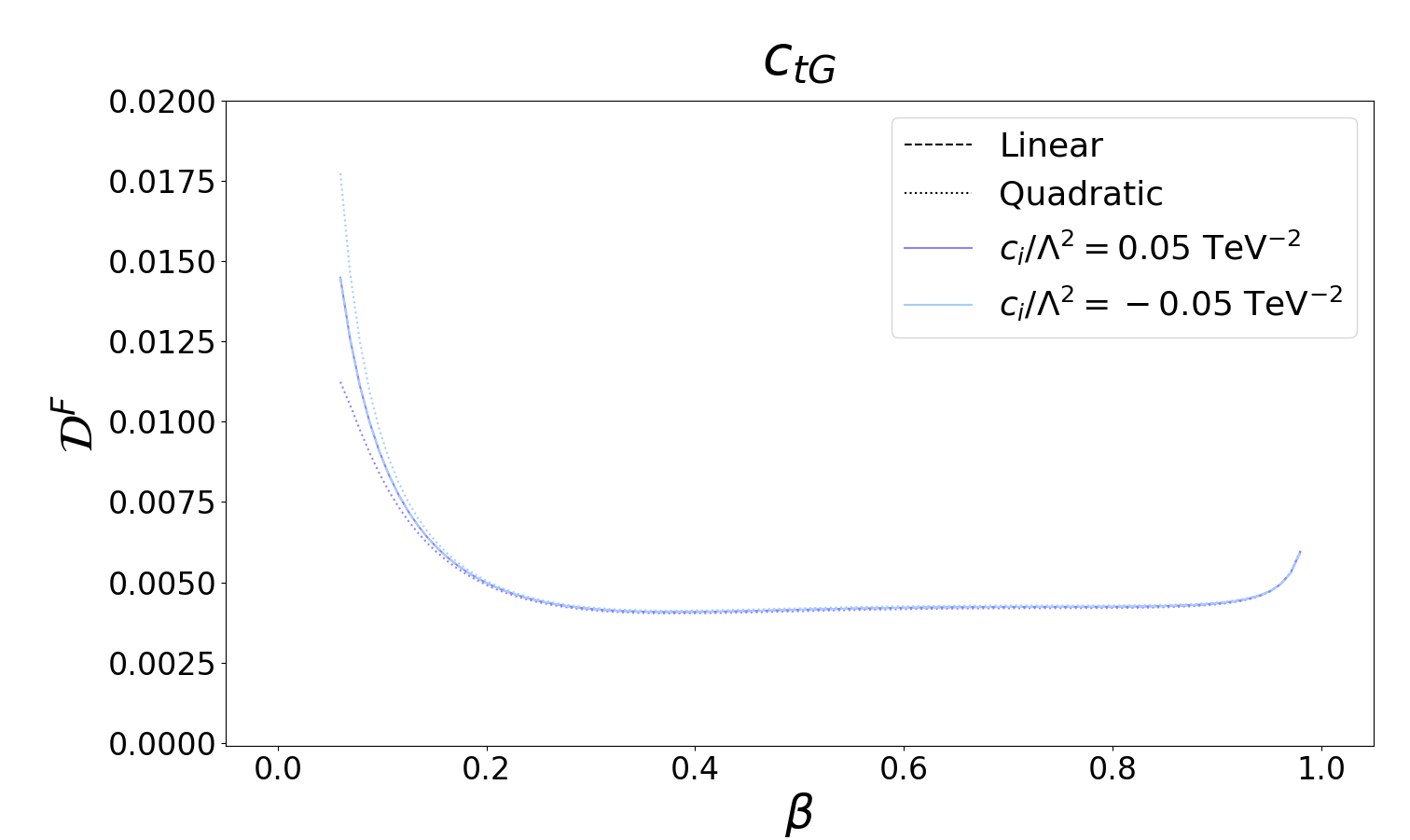}
            \label{fig:fd_ctG_av}
        \end{subfigure}
        \vskip\baselineskip
        \begin{subfigure}[b]{0.475\textwidth}   
            \centering 
            \includegraphics[width=\textwidth]{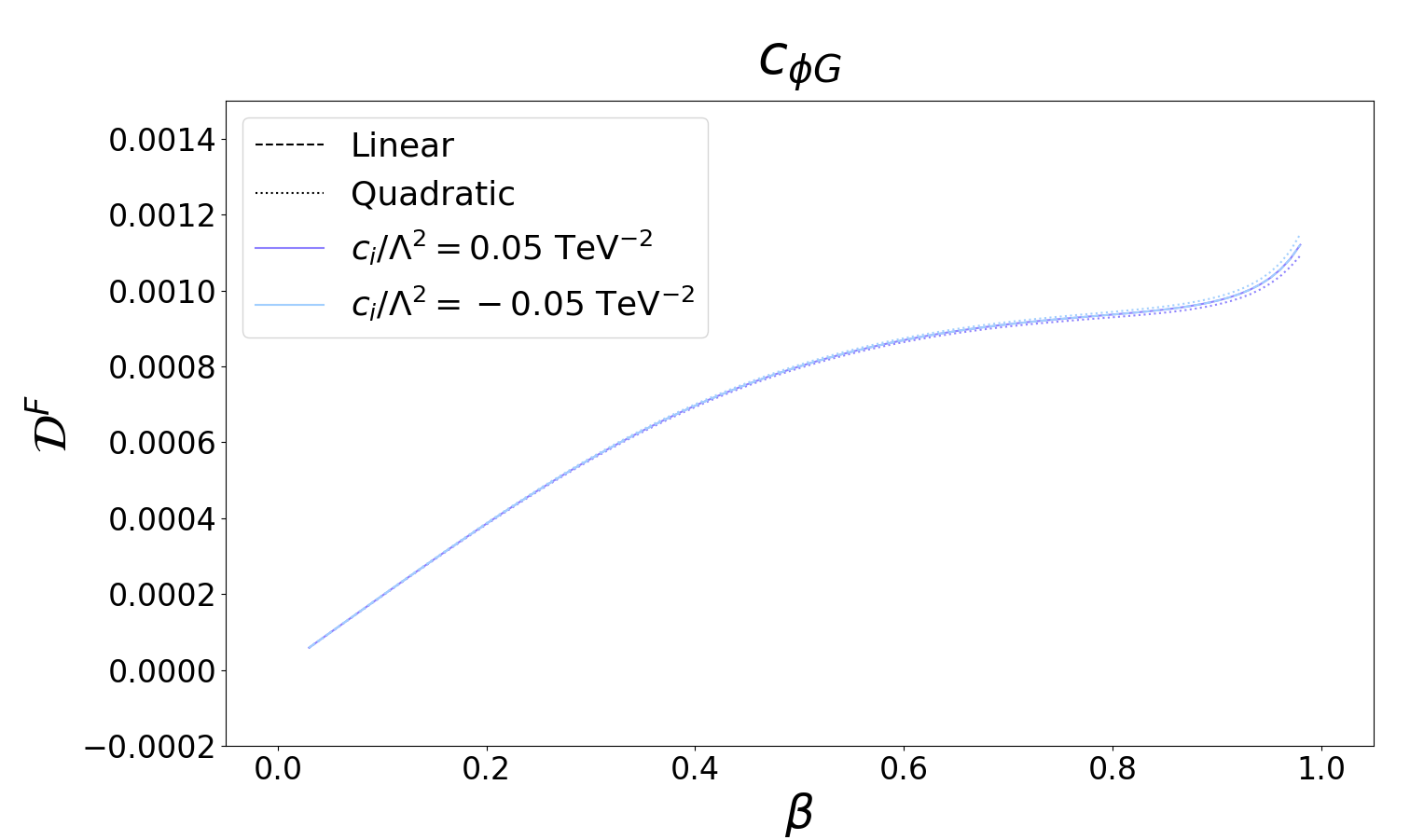}
            \label{fig:fd_cpG_av}
        \end{subfigure}
        \hfill
        \begin{subfigure}[b]{0.475\textwidth}   
            \centering 
            \includegraphics[width=\textwidth]{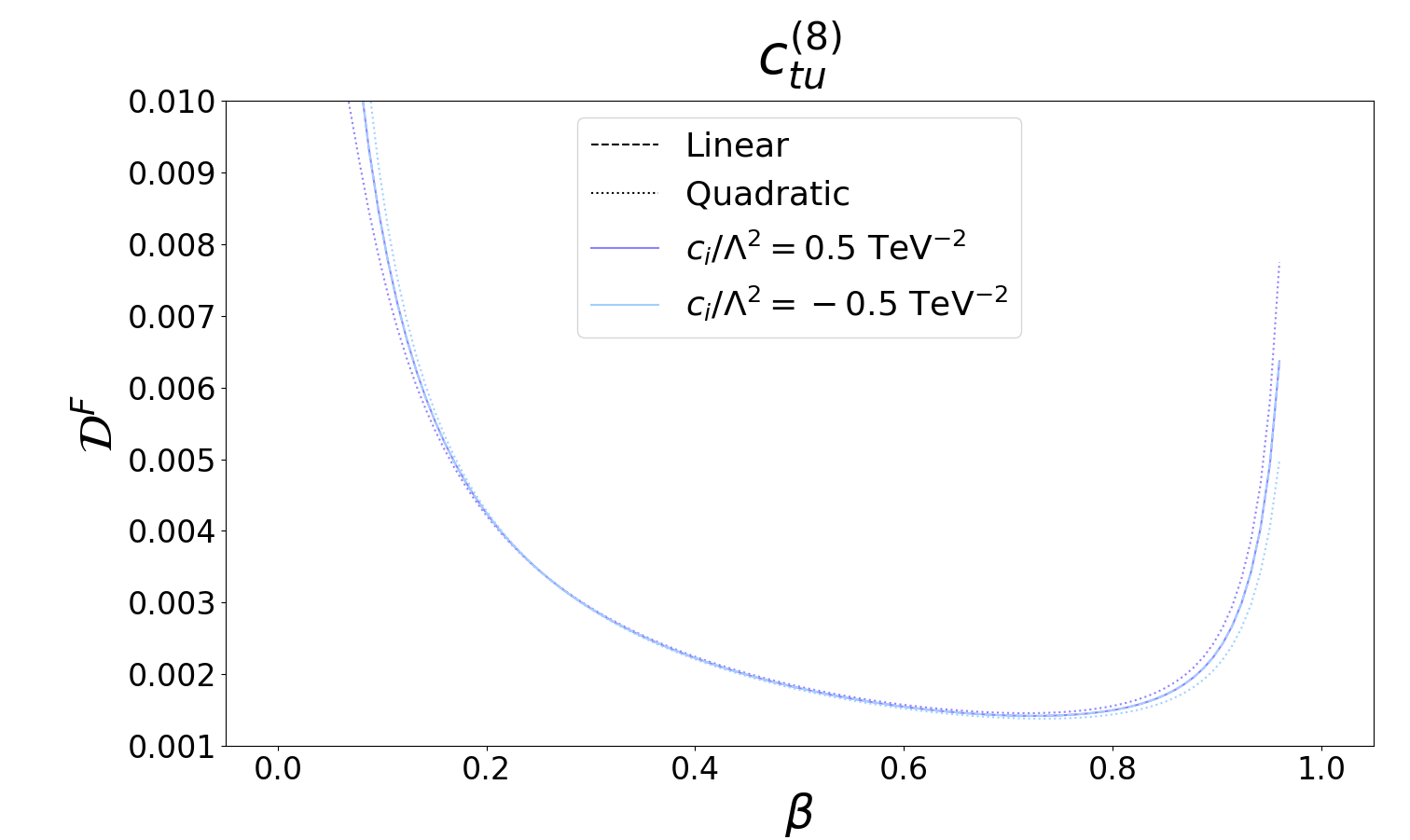}
            \label{fig:fd_ctu8_av}
        \end{subfigure}
        \caption
        {Angular-averaged result for the fidelity distance in the beam basis, for non-zero SMEFT Wilson coefficients as shown, and all other coefficients set to zero. } 
        \label{fig:fidelity}
\end{figure}

In figure~\ref{fig:fidelity}, we show corresponding results to figure~\ref{fig:td}, but for the fidelity distance rather than the trace distance. Similar observations apply, in that divergences at large $\beta$ values are presumably unphysical and can be ignored. The fidelity distance vanishes at threshold, and at linear order in the SMEFT, for $c_{\phi G}$ and $c_G$, as it must do given our above explanation in terms of a SM-like stabiliser state there. Interestingly, numerical values for the fidelity distance are typically larger than for the trace distance. 


Above, we have presented results for a number of quantum information measures, which thus provide differing viewpoints on new physics. To examine this further, we can ask which of these measures is most sensitive to corrections beyond the SM, and adopt the method of ref.~\cite{Fabbrichesi:2025ywl} in order to provide an answer. Following that reference, we define the $\chi^2$ statistic for observable ${\cal O}$ and single non-zero Wilson coefficient $c_I$ as follows:
\begin{equation}
\chi^2({\cal O},c_I)=\left(
\frac{{\cal O}[\rho_{\rm SMEFT}(c_I)]-{\cal O}[\rho_{\rm SM}]}
{\sigma_{\cal O}}
\right)^2~,
\label{chi2def}
\end{equation}
where $\rho_{\rm SM}$ ($\rho_{\rm SMEFT}$) is the (B)SM density matrix, and $\sigma_{\cal O}$ is the uncertainty. For the magic and concurrence, the uncertainty is taken to be that of the SM result, defined as follows. First, we assume a representative fixed uncertainty of 0.2 on the (differential) proton-level $\{C_{ij}\}$, and 0.1 on the $\{B_i^\pm\}$ coefficients\footnote{Our choice of representative uncertainties is motivated by the current error bars in the Fano coefficient measurements of refs.~\cite{CMS:2024hgo,CMS:2025cim}.}. We then evaluate the relevant QI measure ${\cal O}$ -- integrated over a particular bin -- for upper and lower values of the Fano coefficients. The uncertainty $\sigma_{\cal O}$ is then taken to be half the distance between the upper and lower values of ${\cal O}$~\footnote{We note that we only consider experimental uncertainties here. Whilst in principle  theoretical uncertainties e.g. resulting from higher order QCD corrections should be taken into account, they are are expected to be subdominant~\cite{Czakon:2020qbd,Aoude:2025ovu}.}. As discussed in section~\ref{sec:tracedistance}, the trace distance and fidelity distance depend on both the SM and SMEFT density matrices, and also vanish identically in the SM. Equation~(\ref{chi2def}) must then be replaced by
\begin{equation}
\chi^2({\cal O},c_I)=\left(
\frac{{\cal O}[\rho_{\rm SMEFT}(c_I),\rho_{\rm SM}]}
{\sigma_{\cal O}}
\right)^2~.
\label{chi2def2}
\end{equation}
To evaluate the uncertainty, we calculate SM density matrices $\rho^\pm_{\rm SM}$ with the default Fano coefficients plus or minus the uncertainties stated above. We then define the uncertainty to be one half of ${\cal O}[\rho^{+}_{\rm SM},\rho^{-}_{\rm SM}]$. As well as the QI measures described above, we also evaluate the sensitivity of  the Fano coefficients directly, in order to gauge whether there is any particular advantage over such a raw fitting procedure, experimental precedents for which include ref.~\cite{CMS:2019nrx}. To this end, we denote by $FC$ the sum of the $\chi^2$ values for each Fano coefficient individually. 

In a given region in parameter space, the steepness of the rise in $\chi^2$ as the value of a given Wilson coefficient increases in magnitude will differ for different measures ${\cal O}$. One can then rank the measures from most sensitive to least sensitive to new physics, and we present this information for different Wilson coefficients in tables~\ref{tab:ctG}--\ref{tab:ctu8}. In each table, we have integrated our various QI measures ${\cal O}$ in the two-dimensional bins in $(z,m_{t\bar{t}})$ used for the CMS measurement of ref.~\cite{CMS:2024hgo}. For the BSM density matrices, we include up to quadratic SMEFT corrections to the Fano coefficients and consider values for $\bar{c}$ less than 0.005 TeV$^{-2}$. We note that the most sensitive QI measurement very much depends upon both the Wilson coefficient being considered, and the bin. No single measure outperforms all others across the board, which is itself an important lesson to take forwards in applying quantum information ideas to other collider processes. One notable pattern is that the trace distance fares worse for all Wilson coefficients as the invariant mass increases, which can be traced to the fact that the uncertainty on the SM result (as calculated above) increases with $m_{t\bar{t}}$, thereby diminishing the resolving power of this observable. The concurrence, on the other hand, does better as invariant mass increases, in the central bin $|z|<0.4$. For $c_G$, magic (as quantified by the second Stabliser R\'{e}nyi entropy) is the most sensitive QI observable over most of the kinematic range. 

Above, we denoted by $FC$ the $\chi^2$ obtained from considering the Fano coefficients directly, rather than combining them into compound QI measures. Results are shown in tables~\ref{tab:ctG}--\ref{tab:ctu8} alongside the QI measures. We see that, whilst the raw Fano coefficients can indeed outperform the latter in some cases, this is by no means inevitable, such that the relative sensitivity is highly dependent on both the kinematic bin, and the new physics effect being considered. Our results thus agree with previous comments in ref.~\cite{Fabbrichesi:2025ywl}, which stated that fits involving QI measures can be comparable with using the Fano coefficients directly. As also stated in that reference, however, QI measures carry the distinct advantage of an additional conceptual interpretation of new physics constraints, which may ultimately prove useful in elucidating the nature of the new physics itself (i.e. in terms of concrete models). 

\begin{table}[!htb]
\begin{center}
\begin{tabular}{ |l || c | c| }
  \hline			
$  m_{tt}$ [GeV] & $|z| < 0.4 $ & $|z| > 0.4 $\\
  \hline \hline
  $ 2 m_t < m_{tt} < $ 400  & $FC$, $\mathcal{D}^T$, $\mathcal{C}$, $M_2$,  $\mathcal{D}^F$  & $FC$,  $\mathcal{D}^T $,  $ \mathcal{C}$, $M_2$,  $\mathcal{D}^F$  \\
  $400 < m_{tt} < 600 $  & $FC$, $\mathcal{D}^T$, $\mathcal{C}$, $\mathcal{D}^F$, $M_2$ & $FC$,  $\mathcal{C}$, $\mathcal{D}^T$, $M_2$, $\mathcal{D}^F$ \\
    $600 < m_{tt} < 800 $  & $FC$, $ \mathcal{C}$, $\mathcal{D}^T$ , $M_2$, $\mathcal{D}^F$ & $FC$,  $\mathcal{D}^T$, $\mathcal{C}$,  $\mathcal{D}^F$, $M_2$   \\
      $ m_{tt} > 800 $  & $\mathcal{C}$, $M_2$, $\mathcal{D}^F$, $FC$, $\mathcal{D}^T$ & $\mathcal{C}$, $M_2$, $FC$, $\mathcal{D}^F$, $\mathcal{D}^T$  \\
  \hline  
  \hline
  Inclusive & \multicolumn{2}{c|}{$\mathcal{C}$, $FC$, $M_2$, $\mathcal{D}^F$, $\mathcal{D}^T$} \\
\hline
\end{tabular}
  \caption{The relative sensitivity of different quantum information measures to $c_{tG}$, according to the value of their $\chi^2$, and where sensitivity decreases from left to right. For comparison, we also show the relative sensitivity of the Fano coefficients (denoted $FC$) themselves.  }
  \label{tab:ctG}
  \end{center}
\end{table}

\begin{table}[!htb]
\begin{center}
\begin{tabular}{ |l || c | c| }
  \hline			
$  m_{tt}$ [GeV] & $|z| < 0.4 $ & $|z| > 0.4 $\\
  \hline \hline
  $  2m_t < m_{tt} < $ 400  & $M_2$, $FC$,  $\mathcal{D}^F$, $\mathcal{D}^T$, $\mathcal{C}$& $M_2$, $FC$,  $\mathcal{D}^T$,   $\mathcal{D}^F$,  $\mathcal{C}$ \\
  $400 < m_{tt} < 600 $  &  $M_2$,  $\mathcal{D}^F$, $FC$, $\mathcal{D}^T$, $\mathcal{C}$  &  $M_2$, $FC$,   $\mathcal{C}$, $\mathcal{D}^T$, $\mathcal{D}^F$  \\
    $600 < m_{tt} < 800 $  & $M_2$, $\mathcal{D}^F$, $\mathcal{C}$, $FC$, $\mathcal{D}^T$  &  $M_2$, $FC$,  $\mathcal{D}^F$,  $\mathcal{D}^T$,  $\mathcal{C}$ \\
      $ m_{tt} > 800 $  &$\mathcal{C}$, $M_2$, $\mathcal{D}^F$, $FC$,  $\mathcal{D}^T$  & $\mathcal{C}$, $FC$, $M_2$, $\mathcal{D}^F$, $\mathcal{D}^T$   \\
  \hline  
   \hline
  Inclusive & \multicolumn{2}{c|}{$\mathcal{C}$, $M_2$, $\mathcal{D}^F$, $FC$, $\mathcal{D}^T$} \\
\hline
\end{tabular}
  \caption{The relative sensitivity of different quantum information measures to $c_{G}$, according to the value of their $\chi^2$, and where sensitivity decreases from left to right. For comparison, we also show the relative sensitivity of the Fano coefficients (denoted $FC$) themselves.}
  \label{tab:cG}
  \end{center}
\end{table}

\begin{table}[!htb]
\begin{center}
\begin{tabular}{ |l || c | c| }
  \hline			
$  m_{tt}$ [GeV] & $|z| < 0.4 $ & $|z| > 0.4 $\\
  \hline \hline
  $  2m_t < m_{tt} < $ 400  & $FC$, $\mathcal{D}^T$, $\mathcal{C}$,   $M_2$,  $\mathcal{D}^F$, & $FC$, $\mathcal{D}^T$, $\mathcal{C}$, $M_2$, $\mathcal{D}^F$ \\
  $400 < m_{tt} < 600 $  & $FC$, $\mathcal{C}$, $\mathcal{D}^T$,  $M_2$,  $\mathcal{D}^F$   & $FC$,  $\mathcal{D}^T$, $\mathcal{C}$, $\mathcal{D}^F$, $M_2$  \\
    $600 < m_{tt} < 800 $& $FC$, $\mathcal{C}$, $\mathcal{D}^T$, $M_2$,  $\mathcal{D}^F$ & $FC$, $M_2$, $\mathcal{D}^T$, $\mathcal{C}$,  $\mathcal{D}^F$,   \\
      $ m_{tt} > 800 $  &$\mathcal{C}$, $M_2$, $\mathcal{D}^F$, $FC$, $\mathcal{D}^T$ & $\mathcal{C}$, $M_2$, $\mathcal{D}^F$, $FC$, $\mathcal{D}^T$ \\
  \hline 
   \hline
  Inclusive & \multicolumn{2}{c|}{$\mathcal{C}$, $M_2$, $\mathcal{D}^F$, $FC$, $\mathcal{D}^T$} \\
\hline
\end{tabular}
  \caption{The relative sensitivity of different quantum information measures to $c_{tu}^{(8)}$, according to the value of their $\chi^2$, and where sensitivity decreases from left to right. For comparison, we also show the relative sensitivity of the  Fano coefficients (denoted $FC$) themselves.}
  \label{tab:ctu8}
  \end{center}
\end{table}



\section{Conclusions}
\label{sec:conclude}

In this paper, we have examined various quantum information measures as potential probes of new physics effects in top quark pair production at the LHC. Previous studies have examined the role of {\it concurrence} in probing BSM physics~\cite{Aoude:2022imd}, and also shown that the property of {\it magic} (non-stabiliserness) is generically non-zero in the SM~\cite{White:2024nuc}. Other measures, such as the {\it trace distance} and {\it fidelity distance} between two different quantum states, as expressed via their density matrices, have been shown to be useful probes of new physics in other collider processes~\cite{Fabbrichesi:2025ywl}. This paper goes further than this, in examining a range of QI measures as probes of new physics in top quark pair production, where the latter is described by the model-independent Standard Model Effective Field Theory (SMEFT). 

We have first shown how magic -- of high topical interest due to the recent CMS measurement of ref.~\cite{CMS:2025cim} --  gets modified by non-zero SMEFT Wilson coefficients. Corrections are generally non-zero across the whole phase space. Na\"{i}vely, one might think that magic would typically increase, due to the fact that adding extra operators leads to more complicated quantum states in general, that are more likely to move further away from the discrete and special set of stabiliser states in Hilbert space. We find, however, that magic can in fact decrease as well as increase. As for the SM, the profile of magic is dominated by gluon initial states, given the dominance of the gluon PDF at typical LHC partonic centre of mass energies. 

Inspired by refs.~\cite{Aoude:2022imd,Fabbrichesi:2025ywl}, we have also shown results for the angular-averaged trace-distance, fidelity distance and concurrence in the beam basis expanded to both linear and quadratic order in the Wilson coefficients. Corrections are small but non-negligible, and again the EFT expansion appears to be well-behaved, apart from at very high top quark velocities, where the EFT is 
anyway expected to break down. By considering the $\chi^2$ for deviations from the SM, we compared the relative sensitivity of different QI measures in the bins of top quark invariant mass $m_{t\bar{t}}$ and scattering angle $z=\cos\theta$, used in the experimental analysis of ref.~\cite{CMS:2024hgo}. We find that there is no single QI measure that performs best across all of the parameter space, thus suggesting that all can be useful going forwards in efficiently probing new physics effects in different collider processes.
Nor is it true that the QI measures are less sensitive in general than the raw Fano coefficients themselves. Similar remarks in this regard appear in ref.~\cite{Fabbrichesi:2025ywl}, but we here provide quantitative evidence for the first time. We hope that our results provide a useful contribution to the ongoing dialogue between quantum information theory and particle physics, and look forward to further work in this area.

\section*{Acknowledgments}

We thank Eric Madge for fruitful discussions. CDW is supported by the UK Science and Technology Facilities Council
(STFC) Consolidated Grant ST/P000754/1 ``String theory, gauge theory
and duality''. MJW is supported by the Australian Research Council
grants CE200100008 and DP220100007. R.A. is supported by UK Research and Innovation (UKRI) under the UK government’s Horizon Europe Marie Sklodowska-Curie funding guarantee
grant [EP/Z000947/1] and by the STFC grant “Particle Theory at the Higgs Centre”. 
HB acknowledges partial support from the STFC HEP Theory Consolidated grants ST/T000694/1 and ST/X000664/1 and thanks other members of the Cambridge Pheno Working Group for useful discussions.
\appendix
\section{Fano coefficients in the helicity basis}
\label{app:F}

In this appendix we provide expressions for the unnormalised Fano coefficients in the helicity basis as defined in eq.~\ref{rndef}. Whilst these expressions have previously appeared in different sources in literature, we collect them here both  to ensure consistency when including both SM and SMEFT contributions, and for convenience given that we use their analytic properties in interpreting our results.

\subsection{SM Contributions}
\label{app:FanoSM}
We first give expressions for the SM contributions to the unnormalised Fano coefficients. Compared to other presentations in literature (e.g. refs.~\cite{Afik:2022kwm,White:2024nuc}), there is an additional minus sign in the $C_{rk}$ expressions to reflect the difference in our definition of the helicity basis vectors in equation \ref{rndef} to these works.\footnote{The choice of basis vectors in this work was made to reflect that used in ref.~\cite{Aoude:2022imd} from which we take expressions for the SMEFT contributions to the Fano coefficients to be stated in ~\ref{app:Fano}} In order to consistently combine the SM Fano coefficients with the SMEFT contributions (presented in appendix \ref{app:Fano}), each coefficient is additionally multiplied by a factor of $g_s^4$  where $g_s = \sqrt{4 \pi \alpha_s}$ relative to the aforementioned presentations.

In the $gg$ channel, one has 
\begin{subequations}
\begin{align}
    \tilde{A}^{gg,(0)} &=  F_{gg} \big(1 + 2\beta^2(1-z^2)-\beta^4(z^4-2z^2+2)\big),\\
    \tilde{C}^{gg,(0)}_{nn} &= -  F_{gg} \big(1-2\beta^2+\beta^4(z^4-2z^2+2)\big),\\
    \tilde{C}^{gg,(0)}_{kk} &= -  F_{gg} \big(1 - 2 z^2 (1-z^2) \beta^2 - (2 - 2 z^2 + z^4 ) \beta^4 \big),\\
    \tilde{C}^{gg,(0)}_{rr} &=  - F_{gg} \big(1 - (2 - 2 z^2 + z^4 )\beta^2 (2-\beta^2)  \big),\\
    \tilde{C}^{gg,(0)}_{rk} &=  - F_{gg} \, 2 z\, (1-z^2)^{3/2} \beta^2 \sqrt{1-\beta^2},
\end{align}
\end{subequations}
where 
\begin{equation}
F_{gg} =  \frac{g_s^4(7+9\beta^2z^2)}{192(1-\beta^2z^2)^2}~.
\end{equation}
In the $q\bar{q}$-channel the coefficients are
\begin{subequations}
\begin{align}
  \tilde{A}^{q\bar{q},(0)} &= F_{q\bar{q}} \big(2+\beta^2(z^2-1)\big) \\ 
    \tilde{C}^{q\bar{q},(0)}_{nn} &= F_{q\bar{q}} \beta^2\big(z^2-1\big) \\
    \tilde{C}^{q\bar{q},(0)}_{kk} &= F_{q\bar{q}} \big(\beta^2 + z^2 (2 - \beta^2)\big) \\
    \tilde{C}^{q\bar{q},(0)}_{rr} &= F_{q\bar{q}} \big(2-\beta^2 - z^2 (2 - \beta^2)\big) \\
    \tilde{C}^{q\bar{q},(0)}_{rk} &= - 2 z\,F_{q\bar{q}} \sqrt{(1-z^2)(1-\beta^2)}
\end{align}
\end{subequations}
where 
\begin{equation}
F_{q\bar{q}} =  \frac{g_s^4}{18}~.
\end{equation}

As is further explained in ref.~\cite{Aoude:2022imd}, the above results (and indeed those to be presented in appendix~\ref{app:Fano}) assume that the initial-state quark (rather than anti-quark) is travelling in the $+z$ Cartesian direction. In the case that the directions of the quark and anti-quark are interchanged, one must flip $\cos\theta\rightarrow -\cos\theta$, and also take into account an additional overall minus sign in the coefficients $\tilde{C}_{rk}$ and $\tilde{B}_r^\pm$ due to a change of sign of the helicity basis vectors $\vec{n}$ and $\vec{r}$~\cite{CMS:2019nrx}. We emphasise that when summing over different production channels in equation~\ref{Cpp} it is necessary to include both scenarios to reflect the fact that the (anti-) quark could come from either proton.

\subsection{Fano coefficients in the SMEFT}
\label{app:Fano}

We now present the contributions to the unnormalised Fano coefficients entering eq.~(\ref{Rdecomp}) that arise from the various SMEFT operators that contribute to top quark pair production. These results are taken from ref.~\cite{Aoude:2022imd} by one of the present authors. First, one has the following contributions at linear order in the $gg$ channel:
\begin{subequations}
\begin{align}
    \tilde{A}^{gg,(1)} &= \frac{g_s^2}{\Lambda^2} \frac{1}{1-\beta^2 z^2} \bigg[
    \frac{g_s^2 v m_t (9\beta^2z^2+7)}{12\sqrt{2}}c_{tG}
    - \frac{\beta^2m_t^4}{4m_t^2-(1-\beta^2)m_h^2}c_{\varphi G}
    + \frac{9g_s^2\beta^2m_t^2z^2}{8}c_G
    \bigg],\\
   \tilde{C}^{gg,(1)}_{nn} &= \frac{g_s^2}{\Lambda^2} \frac{1}{1-\beta^2 z^2} \bigg[
    \frac{ -7g_s^2v m_t }{12\sqrt{2}}c_{tG}
    - \frac{\beta^2m_t^4}{4m_t^2-(1-\beta^2)m_h^2}c_{\varphi G}
    + \frac{9g_s^2\beta^2m_t^2z^2}{8}c_G
    \bigg],\\
    \tilde{C}^{gg,(1)}_{kk} &= \frac{g_s^2}{\Lambda^2} \frac{1}{1-\beta^2 z^2} \bigg[
        \frac{g_s^2 v m_t \left(9 \beta ^2 z^2+7\right) \left(\beta ^2 \left(z^4-z^2-1\right)+1\right)}{12 \sqrt{2} \left(\beta ^2 z^2-1\right)} c_{tG} 
        \\\notag&\hspace{4cm}
        +\frac{\beta ^2 m_t^4}{4 m_t^2 - \left(1-\beta^2\right) m_h^2} c_{\varphi G} 
        -\frac{9 g_s^2  \beta^2 m_t^2 z^2}{8} c_G
    \bigg],\\
    \tilde{C}^{gg,(1)}_{rr} &= \frac{g_s^2}{\Lambda^2} \frac{1}{1-\beta^2 z^2} \bigg[
        \frac{g_s^2 v m_t \left(-9 \beta ^4 \left(z-z^3\right)^2-7 \beta ^2 \left(z^4-z^2+1\right)+7\right)}{12 \sqrt{2} \left(\beta ^2 z^2-1\right)} c_{tG}
        \\\notag&\hspace{4cm}
        -\frac{\beta ^2 m_t^4}{4 m_t^2-\left(1-\beta^2\right) m_h^2} c_{\varphi G}
        +\frac{9 g_s^2 \beta^2 m_t^2 z^2}{8} c_G
    \bigg], \\
    \tilde{C}^{gg,(1)}_{rk} &= \frac{g_s^2}{\Lambda^2} \frac{1}{1-\beta^2 z^2} \bigg[
        \frac{g_s^2 v m_t \beta^2 z \left(1-z^2\right) \left(9 \beta ^2+\left(\beta ^2-2\right) z^2 \left(9 \beta^2 \left(z^2-1\right)+7\right)-2\right)}{24 \sqrt{2} \sqrt{\left(\beta ^2-1\right) \left(z^2-1\right)} \left(\beta ^2 z^2-1\right)} c_{tG}
        \\\notag&\hspace{4cm}
        + \frac{9 g_s^2 \beta^2 m_t^2 z}{8} \sqrt{\frac{1-z^2}{1-\beta^2}}  c_G
    \bigg].
\end{align}
\end{subequations}
Here $v$ is the Higgs vacuum expectation value, $m_h$ the Higgs boson mass, and $g_s$ the strong coupling constant. 
Corresponding results in the $q\bar{q}$ channel depend upon whether up or down-type quarks are present in the initial state. For up-type quarks interacting via the colour octet 4-fermion operators, non-zero Wilson coefficients appear only through the linear combinations
\begin{equation}
\begin{aligned}
    c_{VV}^{(8),u} &= ( c_{Qq}^{(8,1)} +  c_{Qq}^{(8,3)} + c_{tu}^{(8)} + c_{tq}^{(8)} + c_{Qu}^{(8)})/4,
    \qquad
    &c_{AA}^{(8),u} &= ( c_{Qq}^{(8,1)} +  c_{Qq}^{(8,3)} + c_{tu}^{(8)} - c_{tq}^{(8)} - c_{Qu}^{(8)})/4, \\
    c_{AV}^{(8),u} &= (-c_{Qq}^{(8,1)} -c_{Qq}^{(8,3)} + c_{tu}^{(8)} + c_{tq}^{(8)} - c_{Qu}^{(8)})/4,
    \quad
    &c_{VA}^{(8),u} &= (-c_{Qq}^{(8,1)} -c_{Qq}^{(8,3)} + c_{tu}^{(8)} - c_{tq}^{(8)} + c_{Qu}^{(8)})/4~.
\end{aligned}
\label{cVV}
\end{equation}
For down-type quarks, one must replace $u\rightarrow d$ in the above Wilson coefficients, and reverse the sign of $c_{Qq}^{(8,3)}$. By replacing $(8)\rightarrow (1)$, one obtains similar results for the colour singlet operators.
Results for the Fano coefficients for initial-state up-type quarks are then:
\begin{subequations}
\begin{align}
	\tilde{A}^{q\bar{q},(1)} &= \frac{4 g_s^2 m_t^2}{9 \Lambda^2 (1-\beta^2)} \bigg[
	    \sqrt{2} g_s^2 \frac{v}{m_t} (1-\beta^2) c_{tG}
	    + \left(2-(1-z^2)\beta^2\right) \cVVuO
	    + 2 z \beta \cAAuO
	\bigg],
	\\
	\tilde{C}^{q\bar{q},(1)}_{nn}
	&= - \frac{ g_s^2 m_t^2}{\Lambda^2} \frac{4 \beta^2(1-z^2)}{9(1-\beta^2)} \cVVuO ,\\
	\tilde{C}^{q\bar{q},(1)}_{kk} &= \frac{2 g_s^2 m_t^2}{9 \Lambda^2 (1-\beta^2)} \bigg[
	    2 \sqrt{2} g_s^2 \frac{v}{m_t} (1-\beta^2) z^2 c_{tG}
	    + \big(2 + \beta^2 - (2-\beta^2) (1-2 z^2) \big)\cVVuO
	    + 4 \beta z \cAAuO
	\bigg], \\
	\tilde{C}^{q\bar{q},(1)}_{rr} &= \frac{4 g_s^2 m_t^2 (1-z^2)}{9 \Lambda^2 (1-\beta^2)} \bigg[
	    \sqrt{2} g_s^2 \frac{v}{m_t} (1-\beta^2) c_{tG}
	    + (2-\beta^2) \cVVuO
	\bigg], \\
	\tilde{C}^{q\bar{q},(1)}_{rk} &= -\frac{2 g_s^2 m_t^2}{9 \Lambda^2} \sqrt{\frac{1-z^2}{1-\beta^2}} \bigg[
	    \sqrt{2} g_s^2 \frac{v}{m_t} (2-\beta^2) z c_{tG}
	    + 4 z \cVVuO
	    + 2 \beta \cAAuO
	\bigg],\\
    \tilde{B}_{k}^{\pm,q\bar{q},(1)} &= 4g_s^2\frac{m_t^2}{9\Lambda^2}\frac{1}{1-\beta^2}
    \left(
    \beta(z^2+1)\cAVuO + 2z \cVAuO
    \right),\\
    \tilde{B}_{r}^{\pm,q\bar{q},(1)} &= -4g_s^2\frac{m_t^2}{9\Lambda^2} \sqrt{\frac{1-z^2}{1-\beta^2}}
    \left(\beta z \cAVuO + 2 \cVAuO \right),
\end{align}
\end{subequations}
such that replacing $u\rightarrow d$ yields the appropriate result for down-type quarks. 

As well as the linear results, we have also made use of the following contributions to the Fano coefficients, at quadratic order in the Wilson coefficients, again taken from ref.~\cite{Aoude:2022imd}. For the $gg$ channel one has:
\begin{subequations}
\begin{align}
    \tilde{A}^{gg,(2)} &= \frac{m_t^4}{\Lambda^4} \frac{1}{1-\beta^2} \bigg[ 
    \frac{g_s^4 v^2 (9\beta^4z^4+4\beta^2(3z^2+4)-37)}{24 m_t^2 (\beta^2z^2-1)} c_{tG}^2
    + \frac{24 \beta^2 m_t^4}{((\beta^2-1)m_h^2+4m_t^2)^2} c_{\varphi G}^2
    \\\notag&\hspace{2.25cm}
    +\frac{27 g_s^4 (1-\beta^2z^2)}{4(1-\beta^2)}c_G^2 + \frac{2\sqrt{2}\beta^2 g_s^2 v m_t (z^2-1)}{(\beta^2z^2-1)((\beta^2-1)m_h^2 + 4m_t^2)}c_{tG}c_{\varphi G} + \frac{9 g_s^4 v}{2 m_t \sqrt{2}} c_{tG} c_{G}
    \bigg],\displaybreak[0]\\
    \tilde{C}_{nn}^{gg,(2)}
    &= \frac{m_t^4}{\Lambda^4} \frac{1}{1-\beta^2} \bigg[
        \frac{g_s^4 v^2}{m_t^2} \frac{9\beta^4z^2(z^2-2)+2\beta^2(8z^2-13)+19}{24 (\beta^2z^2-1)}c_{tG}^2
        +\frac{24\beta^2 m_t^4 }{((\beta^2-1)m_h^2 + 4 m_t^2)^2} c_{\varphi G}^2
        \\\notag&\hspace{2.25cm}
        +\frac{27 g_s^4 (\beta^2z^2-1)}{4(\beta^2-1)}c_G^2
        + \frac{2\sqrt{2}\beta^2 g_s^2 v m_t (z^2-1)}{(\beta^2z^2-1)((\beta^2-1)m_h^2+4m_t^2)} c_{tG}c_{\varphi G}
        + \frac{9 g_s^4 v}{2\sqrt{2} m_t} c_{tG}c_{G}
    \bigg],\displaybreak[0]\\
    \tilde{C}_{kk}^{gg,(2)} &= \frac{m_t^4}{\Lambda^4} \frac{1}{1-\beta^2} \bigg[
        \frac{g_s^4 v^2}{24 m_t^2} \frac{1}{(1 - \beta^2 z^2)^2} \bigg(
            9 \beta ^6 z^2 \left(z^4-2\right)
            + \beta ^4 \left(-18 z^6+25 z^4-12 z^2-14\right)
            \\\notag&\hspace{2.25cm}
            + \beta ^2 \left(-28 z^4+81 z^2+12\right)
            - 18 z^2 -19 
        \bigg) c_{tG}^2
        -\frac{24 \beta ^2 m_t^4}{\left(4 m_t^2 - m_h^2 \left(1-\beta ^2\right) \right)^2} c_{\varphi G}^2
        \\\notag&\hspace{2.25cm}
        +\frac{27 g_s^4 \left(1 - \left(2-\beta ^2\right) z^2\right)}{4 \left(1-\beta^2\right)} c_G^2
        -\frac{2 \sqrt{2} g_s^2 v m_t \beta^2 \left(1-z^2\right)}{\left(1-\beta ^2 z^2\right) \left(4 m_t^2 - m_h^2 \left(1-\beta ^2\right)\right)} c_{tG} c_{\varphi G}
        \\\notag&\hspace{2.25cm}
        + \frac{9 g_s^4 v \left(1-\left(2-\beta ^2\right) z^2\right)}{2 \sqrt{2} m_t \left(1-\beta ^2 z^2\right)} c_{tG} c_G
    \bigg], \displaybreak[0]\\
    \tilde{C}_{rr}^{gg,(2)} &= \frac{m_t^4}{\Lambda^4} \frac{1}{1-\beta^2} \bigg[
        \frac{g_s^4 v^2}{24 m_t^2} \frac{1}{(1 - \beta^2 z^2)^2} \bigg(
            -9 \beta ^6 z^2 \left(z^4-2 z^2+2\right)
            + \beta ^4 \left(18 z^6-57 z^4+52 z^2-14\right)
            \\\notag&\hspace{2.25cm}
            + \beta ^2 \left(28 z^4-57 z^2+58\right)
            +18 z^2-37
        \bigg) c_{tG}^2
        + \frac{24 \beta^2 m_t^4}{\left(4 m_t^2 - m_h^2 \left(1-\beta ^2\right)\right)^2} c_{\varphi G}^2
        \\\notag&\hspace{2.25cm}
        - \frac{27 g_s^4 \left(1-\left(2-\beta ^2\right) z^2\right)}{4 \left(1-\beta ^2\right)} c_G^2
        + \frac{2 \sqrt{2} g_s^2 v m_t \beta^2 \left(1-z^2\right)}{\left(1-\beta ^2 z^2\right) \left(4 m_t^2 - m_h^2 \left(1-\beta ^2\right)\right)} c_{tG} c_{\varphi G}
        \\\notag&\hspace{2.25cm}
        - \frac{9 g_s^4 v \left(1-\left(2-\beta ^2\right) z^2\right)}{2 \sqrt{2} m_t \left(1-\beta ^2 z^2\right)} c_{tG} c_{G}
    \bigg],\displaybreak[0] \\
    \tilde{C}_{rk}^{gg,(2)} &= \frac{m_t^4}{\Lambda^4} \frac{1}{\sqrt{1-\beta^2}} \bigg[
        \frac{g_s^4 v^2}{192 m_t^2} \frac{1}{(1 - \beta^2 z^2)^2} \bigg(
            - 144 \beta^4 z^3 \left(1-z^2\right)^{3/2}
            + 16 \beta^2 z \sqrt{1-z^2} \left(14 z^2-23\right) 
            \\\notag&\hspace{2.25cm}
            + 144 z \sqrt{1-z^2} 
        \bigg) c_{tG}^2
        + \frac{27}{2} g_s^4 z\, \frac{\sqrt{1-z^2}}{1-\beta^2} c_{G}^2
        - \frac{2 \sqrt{2} g_s^2 v m_t \beta^2 z \sqrt{1-z^2}}{\left(1-\beta ^2 z^2\right) \left(4 m_t^2 - m_h^2 \left(1-\beta ^2\right)\right)} c_{tG} c_{\varphi G}
        \\\notag&\hspace{2.25cm}
        +\frac{9 g_s^4 v z \sqrt{1-z^2}}{\sqrt{2} m_t \left(1-\beta ^2 z^2\right)} c_{tG} c_G
    \bigg].
\end{align}
\end{subequations}
For the up-type contribution to the $q\bar{q}$ channel one has:
\begin{subequations}
\begin{align}
    \tilde{A}^{q\bar{q},(2)} &= \frac{4 m_t^4}{9(1-\beta^2)^2\Lambda^4} \bigg[
        \frac{g_s^4 v^2}{m_t^2} \left(1-\beta ^2\right)  \left(2-\beta ^2 \left(z^2+1\right)\right) c_{tG}^2
        +\frac{g_s^2 v}{m_t} 4 \sqrt{2} \left(1-\beta ^2\right) (\cVVuO + \beta z \cAAuO ) c_{tG}
        \\\notag&\phantom{=}\ 
        + \beta ^2 \bigg(9 (\cAAuS)^2 \left(z^2+1\right)
        +2 (\cAAuO)^2 \left(z^2+1\right)
        +z^2 \left(9 (\cAVuS)^2+2 (\cAVuO)^2+9 (\cVAuS)^2+2 (\cVAuO)^2
        \right.\\\notag&\phantom{=}\ \left.+9 (\cVVuS)^2+2 (\cVVuO)^2\right)
        +9 (\cAVuS)^2+2 (\cAVuO)^2-9 (\cVAuS)^2-2 (\cVAuO)^2 -9 (\cVVuS)^2 -2 (\cVVuO)^2\bigg)
        \\\notag&\phantom{=}\
        +4 \beta  z (9 \cAAuS \cVVuS+2 \cAAuO \cVVuO+9 \cAVuS \cVAuS+2 \cAVuO \cVAuO)
        \\\notag&\phantom{=}\
        +18 (\cVAuS)^2+4 (\cVAuO)^2+18 (\cVVuS)^2+4 (\cVVuO)^2
    \bigg] ,
    \displaybreak[0]\\
    \tilde{C}_{nn}^{q\bar{q},(2)} &=  \frac{4 m_t^4}{9 \Lambda^4} \frac{\beta^2 ( 1-z^2)}{(1-\beta^2)^2}\bigg[
        \frac{g_s^4 v^2}{m_t^2} \left(1-\beta ^2\right) c_{tG}^2
        + 9 (\cAAuS)^2+2 (\cAAuO)^2+9 (\cAVuS)^2+2 (\cAVuO)^2-9 (\cVAuS)^2
        \\\notag&\phantom{=}\
        -2 (\cVAuO)^2-9 (\cVVuS)^2-2 (\cVVuO)^2
    \bigg] ,
    \displaybreak[0]\\
    \tilde{C}_{kk}^{q\bar{q},(2)} &=  \frac{4 m_t^4}{9 (1-\beta^2)^2\Lambda^4} \bigg[
        \frac{g_s^4 v^2}{m_t^2} \left(1-\beta ^2\right) \left( z^2(2-\beta^2)-\beta^2 \right) c_{tG}^2
        + \frac{g_s^2 v}{m_t} 4 \sqrt{2} \left(1-\beta ^2\right) z \left(\beta \cAAuO + z \cVVuO\right) c_{tG}
        \\\notag&\phantom{=}\
        + \beta ^2 \left(
            9 (\cAAuS)^2 \left(z^2+1\right)+2 (\cAAuO)^2 \left(z^2+1\right)
            +z^2 \left(9 (\cAVuS)^2+2 (\cAVuO)^2-9 (\cVAuS)^2-2 (\cVAuO)^2
            \right.\right.\\\notag&\phantom{=}\ \left.\left.
            -9 (\cVVuS)^2-2 (\cVVuO)^2\right)
            +9 (\cAVuS)^2+2 (\cAVuO)^2+9 (\cVAuS)^2+2 (\cVAuO)^2+9 (\cVVuS)^2+2 (\cVVuO)^2
        \right)
        \\\notag&\phantom{=}\
        +4 \beta  z (9 \cAAuS \cVVuS+2 \cAAuO \cVVuO+9 \cAVuS \cVAuS+2 \cAVuO \cVAuO)
        \\\notag&\phantom{=}\
        +2 z^2 \left(9 (\cVAuS)^2+2 (\cVAuO)^2+9 (\cVVuS)^2+2 (\cVVuO)^2\right)
    \bigg] ,
    \displaybreak[0]\\
    \tilde{C}_{rr}^{q\bar{q},(2)} &=  \frac{4 m_t^4}{9 \Lambda^4} \frac{1-z^2}{(1-\beta^2)^2} \bigg[
        \frac{g_s^4 v^2}{m_t^2} ( 2 - 3 \beta^2 + \beta^4 ) c_{tG}^2
        + \frac{g_s^2 v}{m_t} 4 \sqrt{2} \left(1-\beta^2\right) \cVVuO c_{tG}
        \\\notag&\phantom{=}\
        -\beta ^2 \left( 
            9 (\cAAuS)^2+2 (\cAAuO)^2
            +9 (\cAVuS)^2+2 (\cAVuO)^2+2 (\cVAuO)^2+9 (\cVVuS)^2+2 (\cVVuO)^2
        \right)
        \\\notag&\phantom{=}\
        -9 \left(\beta ^2-2\right) (\cVAuS)^2
        +4 (\cVAuO)^2+18 (\cVVuS)^2+4 (\cVVuO)^2
    \bigg] ,
    \displaybreak[0]\\
        \tilde{C}_{rk}^{q\bar{q},(2)} &= -\frac{8 m_t^4}{9 (1-\beta^2) \Lambda^4} \sqrt{\frac{1-z^2}{1-\beta^2}}\,\bigg[
            \frac{g_s^4 v^2}{m_t^2} z (1-\beta^2) c_{tG}^2
            + \frac{g_s^2 v}{m_t} \sqrt{2} \left( z (2-\beta^2) \cVVuO + \beta \cAAuO \right) c_{tG}
            \\\notag&\phantom{=}\
            + \beta  \left(9 \cAAuS \cVVuS+2 \cAAuO \cVVuO+9 \cAVuS \cVAuS+2 \cAVuO \cVAuO\right)
            \\\notag&\phantom{=}\
            +z \left(9 (\cVAuS)^2+2 (\cVAuO)^2+9 (\cVVuS)^2+2 (\cVVuO)^2\right)
        \bigg] ,
\end{align}
\end{subequations}
where the down-type contribution is obtained by replacing $u\rightarrow d$ as before. 

\section{Angular-averaged Fano coefficients}
\label{app:ang}
We finally present expressions for the Fano coefficients in the \textit{beam} basis (as opposed to the top-direction dependent helicity basis) averaged over solid angle as per equation~\ref{Rav}. These results are again taken from ref.~\cite{Aoude:2022imd}. 

In the SM, one has~\cite{Afik:2020onf}
\begin{subequations}
\begin{align}
    \tilde{A}^{gg,(0)}_\Omega &= \frac{g_s^4}{192} \left[  
        -59 + 31 \beta^2
        + 2 ( 33 - 18 \beta^2 + \beta^4 ) \frac{\artanh\beta}{\beta}
    \right] , \displaybreak[0]\\
    \tilde{C}^{gg,(0)}_{z,\Omega} &= \frac{g_s^4}{2880 \beta^4}\bigg[
        879 \beta^6 - 3413 \beta^4 + 4450 \beta^2 - 2940 
        + 4 ( 72 \beta^4 - 745 \beta^2 + 735 ) \sqrt{1-\beta^2}
        \\\notag&\hspace{2cm}
        + 30 \left( \beta^8 -53 \beta^6 + 151 \beta^4 - 181 \beta^2 + 98  - 2 ( 17 \beta^4 -66\beta^2 + 49 ) \sqrt{1-\beta^2} \right) \frac{\artanh\beta}{\beta}
    \bigg],\displaybreak[0]\\
    \tilde{C}^{gg,(0)}_{\perp,\Omega} &=  \frac{g_s^4}{2880 \beta^4}\bigg[
        -207 \beta^6 + 1024 \beta^4 - 2225 \beta^2 + 1470
        - 2 ( 72 \beta^4 - 745 \beta^2 + 735 ) \sqrt{1-\beta^2}
        \\\notag&\hspace{2cm}
        + 15 \left( 33 \beta^6 - 116 \beta^4 + 181 \beta^2 - 98  + 2 ( 17 \beta^4 -66\beta^2 + 49 ) \sqrt{1-\beta^2} \right) \frac{\artanh \beta}{\beta}
    \bigg],
\end{align}
\end{subequations}
in the $gg$ channel, whereas in the $q\bar{q}$ channel one has
\begin{align}
    \tilde{A}^{q\bar{q},(0)}_\Omega &=  \frac{g_s^4(3-\beta^2)}{27}\,,  &\qquad
    \tilde{C}^{q\bar{q},(0)}_{z,\Omega} &= \frac{g_s^4(11 - 3 \beta^2 + 4 \sqrt{1-\beta^2})}{135} \,,&\qquad
    \tilde{C}^{q\bar{q},(0)}_{\perp,\Omega} &=  \frac{g_s^4(2 - \beta^2 - 2 \sqrt{1-\beta^2})}{135}\,.
\end{align}

The contributions from the SMEFT operators at linear order are given by
\begin{subequations}
\begin{align}
    \tilde{A}^{gg,(1)}_\Omega &=  -\frac{g_s^2 m_t^2}{\Lambda^2} \left[
        \frac{g_s^2 v}{m_t} \frac{9\beta - 16 \artanh \beta}{12 \sqrt{2} \beta} c_{tG}
        + \frac{m_t^2 \beta \artanh\beta}{4 m_t^2 - m_h^2 (1-\beta^2)} c_{\varphi G}
        + 9 g_s^2 \frac{\beta-\artanh \beta}{8 \beta} c_{G}
    \right],  \displaybreak[0]\\
    \tilde{C}^{gg,(1)}_{z,\Omega} &= - \frac{m_t^2}{\Lambda^2} \bigg[ 
        \frac{\sqrt{2}  g_s^4 v}{360\, m_t} \bigg( 
            \frac{1470 -1865 \beta^2 + 529 \beta^4 - 72 \beta^6}{\beta^4 \sqrt{1-\beta^2}}
            + \frac{1470 - 1130 \beta^2 + 264 \beta^4}{\beta^4}
            \\\notag&\hspace{1.5cm}
            + 15 \frac{9 \beta^6 - 68 \beta^4 + 157 \beta^2 -98 + \left(\beta^6 -27 \beta^4 + 108 \beta^2 -98\right)\sqrt{1-\beta^2}}{\beta^4 \sqrt{1-\beta^2}}  \frac{\artanh{\beta}}{\beta}
        \bigg) c_{tG} 
        \\\notag&\hspace{1.5cm}
        + \frac{g_s^2 m_t^2}{\beta} \frac{2 \beta - (2-\beta^2) \artanh\beta}{4 m_t^2 - m_h^2 (1-\beta^2)} c_{\varphi G}
        \\\notag&\hspace{1.5cm}
        + \frac{3 g_s^4}{8 \beta^3} \frac{4 \beta^3 - 6 \beta + (\beta^3 - 6 \beta)\sqrt{1-\beta^2} + \left(6 - 6 \beta^2 + 3 (2-\beta^2)\sqrt{1-\beta^2}\right)\artanh\beta}{\sqrt{1-\beta^2}} c_G
    \bigg],\displaybreak[0]\\
    \tilde{C}^{gg,(1)}_{\perp,\Omega} &= \frac{m_t^2}{\Lambda^2} \bigg[
        \frac{\sqrt{2} g_s^4 v}{720 m_t \beta^4 (1-\beta^2)} \bigg(
            111 \beta^6 + 1019 \beta^4 -2600 \beta^2 + 1470 
            \\\notag&\hspace{1.5cm}
            - (72 \beta^6 - 529 \beta^4 + 1865 \beta^2 - 1470) \sqrt{1-\beta^2} 
            - 15 \left(1-\beta^2\right)\left( \beta^6 + 23 \beta^4 -108 \beta^2 + 98 \right) \frac{\artanh\beta}{\beta}
            \\\notag&\hspace{1.5cm}
            - 15 \left( 9 \beta^4 - 59 \beta^2 + 98 \right) \left(1-\beta^2\right)^{3/2} \frac{\artanh\beta}{\beta}
        \bigg) c_{tG}
        + \frac{g_s^2 m_t^2 \left(\beta - \artanh\beta\right)}{\beta \left(4 m_t^2 - m_h^2 (1-\beta^2)\right)} c_{\varphi G}
        \\\notag&\hspace{1.5cm}
        + \frac{3 g_s^4}{8} \frac{2 \beta^3 - 3 \beta - (\beta^3 +3\beta)\sqrt{1-\beta^2} + 3 \left( 1-\beta^2 +\sqrt{1-\beta^2} \right)\artanh\beta}{\beta^3 \sqrt{1-\beta^2}} c_G 
    \bigg],
\end{align}
\end{subequations}
\begin{subequations}
\begin{align}
    \tilde{A}^{q\bar{q},(1)}_\Omega &= \frac{4 g_s^2 m_t^2}{27 \Lambda^2} \left[ 3 \sqrt{2}\, \frac{g_s^2 v}{m_t}\, c_{tG} + 2 \frac{3-\beta^2}{1-\beta^2} \,\cVVuO\, \right], \displaybreak[0]\\
    \tilde{C}^{q\bar{q},(1)}_{z,\Omega} &= \frac{8 g_s^2 m_t^2}{135 \Lambda^2} \left[ 
        \frac{g_s^2 v}{m_t} \frac{2 \beta^2-4+11 \sqrt{1-\beta^2}}{\sqrt{2-2 \beta^2}}\, c_{tG}+\frac{11-3 \beta^2-4 \sqrt{1-\beta^2}}{1-\beta^2} \,\cVVuO
    \right], \displaybreak[0]\\
    \tilde{C}^{q\bar{q},(1)}_{\perp,\Omega} &= \frac{8 g_s^2 m_t^2}{135 \Lambda ^2} \left[
        \frac{g_s^2 v}{m_t} \frac{2-\beta ^2+2 \sqrt{1-\beta ^2}}{\sqrt{2-2 \beta ^2}} c_{tG}
        + \frac{2-\beta ^2+2 \sqrt{1-\beta ^2}}{1-\beta ^2} \cVVuO
    \right], \displaybreak[0]\\
    B_{z,\Omega}^{q\bar{q},(1)} &= \frac{8 g_s^2 m_t^2}{27 \Lambda ^2} \frac{1-2 \sqrt{1-\beta ^2}}{1-\beta ^2} \cVAuO
\end{align}
\end{subequations}
and the dimension-six squared corrections are
\begin{subequations}
\begin{align}
    \tilde{A}^{gg,(2)}_\Omega &= \frac{m_t^4}{\Lambda^4} \frac{1}{1-\beta^2} \bigg[
        \frac{g_s^4 v^2}{m_t^2} \frac{3 \beta (7+\beta^2) + 16 (1-\beta^2) \artanh\beta}{24 \beta} c_{tG}^2
        + \frac{24 m_t^4 \beta^2}{\left(4 m_t^2 - m_h^2 (1-\beta^2)\right)^2} c_{\varphi G}^2
        \\\notag&\hspace{2.5cm}
        + \frac{9 g_s^4}{4} \frac{3-\beta^2}{1-\beta^2} c_G^2
        + 2 \sqrt{2} g_s^2 m_t v \frac{\beta - (1-\beta^2)\artanh\beta}{\beta \left(4 m_t^2 - m_h^2 (1-\beta^2)\right)} c_{tG} c_{\varphi G}
        + \frac{9 g_s^4 v}{2 \sqrt{2} m_t} c_{tG} c_G  
    \bigg], \displaybreak[0]\\
    \tilde{C}^{gg,(2)}_{z,\Omega} &= \frac{m_t^4}{\Lambda^4} \frac{1}{1-\beta^2} \bigg[
        \frac{g_s^4 v^2}{360 m_t^2 \beta^4} \bigg(
            429 \beta^6 - 2773 \beta^4 + 7100 \beta^2 -4800
            - \left(288 \beta^4 - 4700 \beta^2 + 4800 \right) \sqrt{1-\beta ^2}
            \\\notag&\phantom{=}\
            + 30 \left(\beta^8 - 12 \beta^6 +141 \beta^4 -290 \beta^2 + 160 + 10 \left(5 \beta^4-21 \beta^2+16\right) \sqrt{1-\beta ^2} \right) \frac{\artanh\beta}{\beta}
        \bigg) c_{tG}^2
        \\\notag&\phantom{=}\
        + \frac{8 m_t^4 \beta ^2}{\left(4 m_t^2 - m_h^2(1-\beta^2)\right)^2} c_{\varphi G}^2
        + \frac{9 g_s^4}{20} \frac{\beta^2-7+8\sqrt{1-\beta^2}}{1-\beta^2} c_G^2
        -\frac{2 \sqrt{2} g_s^2 m_t v}{3 \beta ^3 \sqrt{1-\beta ^2}} \bigg(
            \frac{4 \beta^5 - 10 \beta^3 + 6 \beta}{4 m_t^2 - (1-\beta^2) m_h^2}
            \\\notag&\phantom{=}\
            + \frac{ 6 \beta -7\beta^3 }{4 m_t^2 - (1-\beta^2) m_h^2} \sqrt{1-\beta ^2}
            - 3 \frac{2 (1-\beta^2)^2 + (\beta^4 -3 \beta^2 + 2) \sqrt{1-\beta ^2}}{4 m_t^2 - (1-\beta^2) m_h^2} \artanh\beta
        \bigg) c_{tG} c_{\varphi G}
        \\\notag&\phantom{=}\
        + \frac{3 \sqrt{2} g_s^4 v}{4 m_t \beta^5 \sqrt{1-\beta^2}} \bigg(
            8 \beta^5 -20 \beta^3 + 12 \beta
            + ( \beta^5 - 14 \beta^3 + 12 \beta) \sqrt{1-\beta^2} 
            \\\notag&\phantom{=}\
            - 6 (1-\beta^2) \left( 2 (1-\beta^2) + (2-\beta^2) \sqrt{1-\beta^2} \right) \artanh\beta
        \bigg) c_{tG} c_G
    \bigg],  
    \displaybreak[0]\\
    \tilde{C}^{gg,(2)}_{\perp,\Omega} &= \frac{m_t^4}{\Lambda^4} \frac{1}{1-\beta^2} \bigg[
        \frac{g_s^4 v^2}{360 m_t^2 \beta^4} \bigg(
            48 \beta^6+ 1094 \beta^4 - 3550 \beta^2 + 2400
            + \left(144 \beta ^4-2350 \beta ^2+2400 \right) \sqrt{1-\beta ^2}
            \\\notag&\phantom{=}\
            + 15 \left( \beta^8 + 28 \beta^6 - 159 \beta^4 + 290 \beta^2 - 160 - 10 \left(5 \beta^4 - 21 \beta^2 + 16\right) \sqrt{1-\beta ^2}\right) \frac{\artanh\beta}{\beta}
        \bigg)c_{tG}^2
        \\\notag&\phantom{=}\
        + \frac{8 m_t^4 \beta ^2}{\left(4 m_t^2 - m_h^2(1-\beta^2)\right)^2} c_{\varphi G}^2
        - \frac{9 g_s^4}{20} \frac{3 \beta^2-11+4\sqrt{1-\beta^2}}{1-\beta^2} c_G^2
        \\\notag&\phantom{=}\
        - \frac{2\sqrt{2} g_s^2 m_t v}{3 \beta^2 \left(4 m_t^2 - m_h^2(1-\beta^2)\right)} \left(1 + \sqrt{1-\beta^2}\right) \left(
            2 \beta^2 - 3 + 3 (1-\beta^2) \frac{\artanh\beta}{\beta}
        \right) c_{tG} c_{\varphi G}
        \\\notag&\phantom{=}\
        + \frac{3 \sqrt{2} g_s^4 v}{4 m_t \beta^4} \bigg(
            \beta^4 + 7 \beta^2 - 6 
            + (4\beta^2-6)\sqrt{1-\beta^2}
            + 3 \left(
                2 - 3 \beta^2 + \beta^4
                + 2 (1-\beta^2)^{3/2}
            \right) \frac{\artanh\beta}{\beta}
        \bigg) c_{tG} c_G
    \bigg],
\end{align}
\end{subequations}
\begin{subequations}
\begin{align}
    \tilde{A}^{q\bar{q},(2)}_\Omega &= \frac{8 m_t^4}{27 \Lambda^4} \bigg[ 
        \frac{g_s^4 v^2}{m_t^2} \frac{3 - 2 \beta^2}{1-\beta^2}\, c_{tG}^2
        + 6 \sqrt{2} \, \frac{g_s^2 v}{m_t} \frac{1}{1-\beta^2}\, c_{tG} \cVVuO \\
        &\hspace{2cm}+ \frac{3-\beta^2}{(1-\beta^2)^2} \left( 9 (\cVVuS)^2 + 9 (\cVAuS)^2 + 2 (\cVVuO)^2 + 2 (\cVAuO)^2\right) \notag\\
        &\hspace{2cm}+ \frac{2\,\beta^2}{(1-\beta^2)^2} \left( 9 (\cAAuS)^2 + 9 (\cAVuS)^2 + 2 (\cAAuO)^2 + 2 (\cAVuO)^2\right)
    \bigg], \notag \displaybreak[0]\\
    \tilde{C}^{q\bar{q},(2)}_{z,\Omega} &= \frac{8 m_t^4}{135 \Lambda ^4} \frac{1}{\left(1-\beta ^2\right)^2} \bigg[
        \frac{g_s^4 v^2 }{m_t^2} (1-\beta^2) \left(11-8 \beta^2-4 \sqrt{1-\beta ^2}\right) c_{tG}^2 
        \\&\hspace{3.25cm}
        + 2 \sqrt{2} \frac{g_s^2 v}{m_t} \left(11 \left(1-\beta ^2\right)+\left(2 \beta ^2-4\right) \sqrt{1-\beta ^2}\right) c_{tG} \cVVuO
        \notag\\&\hspace{3.25cm}
        + \left(11 -3 \beta ^2-4 \sqrt{1-\beta ^2}\right) \left( 9 (\cVVuS)^2 + 9 (\cVAuS)^2 + 2 (\cVVuO)^2 + 2 (\cVAuO)^2 \right)
    \bigg], \notag \displaybreak[0]\\
    \tilde{C}^{q\bar{q},(2)}_{\perp,\Omega} &= \frac{8 m_t^4}{135 \Lambda ^4} \frac{1}{\left(1-\beta ^2\right)^2} \bigg[
        \frac{g_s^4 v^2 }{m_t^2} (1-\beta^2) \left(2-\beta ^2+2 \sqrt{1-\beta ^2}\right) c_{tG}^2
        \\&\hspace{3.25cm}
        + 2 \sqrt{2} \frac{g_s^2 v}{m_t} \left(2 \left(1-\beta ^2\right)+\left(2-\beta ^2\right) \sqrt{1-\beta ^2}\right) c_{tG} \cVVuO
        \notag\\&\hspace{3.25cm}
        + 5 \beta^2 \left( 9 (\cAAuS)^2 + 9 (\cAVuS)^2 + 2 (\cAAuO)^2 + 2 (\cAVuO)^2 \right)
        \notag\\&\hspace{3.25cm}
        + \left(2-\beta^2 + 2 \sqrt{1-\beta^2}\right) \left( 9 (\cVVuS)^2 + 9 (\cVAuS)^2 + 2 (\cVVuO)^2 + 2 (\cVAuO)^2 \right)
    \bigg]. \notag
\end{align}
\end{subequations}

\providecommand{\href}[2]{#2}\begingroup\raggedright\endgroup

\end{document}